\documentclass{aa}

\input{epsf.sty}
\def\plotone#1{\centering \leavevmode 
\epsfxsize=\columnwidth \epsfbox{#1}}

\def\p{\partial}
\def\varp{\varphi}
\def\vart{\vartheta}
\def\rund#1{\left(#1\right)}
\def\eck#1{\left\lbrack#1\right\rbrack}

\def\mo{M_\odot}
\def\msol{\mbox{M}_\odot}
\def\dt{\frac{d\ln R_s}{dt}}
\def\pt{\rund{\frac{\p\ln R_s}{\p t}}}
\def\pto{\rund{\frac{\p\ln R}{\p t}}}

\def\dm{\dot{M}_s} 
\def\dmm{\frac{\dot{M}_s}{M_s}} 

\def\irr{{\small{\rm irr}}}
\def\eff{{\small{\rm eff}}}
\def\nuc{{\small{\rm nuc}}}
\def\rad{{\small{\rm rad}}}
\def\ml{{\small{\rm ml}}}
\def\max{{\small{\rm max}}}
\def\conv{{\small{\rm conv}}}
\def\core{{\small{\rm core}}}
\def\iint{\int^{2\pi}_0\int^\pi_0}

\def\be{\begin{equation}}
\def\ee{\end{equation}}
\def\bea{\begin{eqnarray}}
\def\eea{\end{eqnarray}}
\newcommand{\rev}{}

\begin{document}

\thesaurus{02.01.2 
 08.02.1 
 08.05.3 
 08.09.3 
 08.14.2 
 13.25.5 }

\title{Irradiation and mass transfer in low-mass compact binaries}
\author{H.~Ritter\inst{1}, Z.-Y.~Zhang\inst{2,3} and U. Kolb\inst{4}}

\offprints{hsr@mpa-garching.mpg.de}

\institute{Max-Planck-Institut f\"ur Astrophysik, 
Karl-Schwarzschild-Str.\ 1, D-85740 Garching, Germany \and 
National Astronomical Observatories, Chinese Academy of Sciences,
Beijing 100012, P.R. China \and
Department of Physics \& Astronomy, University of Leicester, 
Leicester LE1 7RH, UK \and
Department of Physics \& Astronomy, The Open University,
Milton Keynes MK7 6AA, UK}
\date{revised version, accepted 23 May 2000} 

\authorrunning{H.~Ritter, Z.-Y.~Zhang, U.~Kolb}
\titlerunning{Irradiation and mass transfer}
\maketitle
\sloppy


\begin{abstract}
This paper studies the reaction of low-mass stars to anisotropic
irradiation and its implications for the long-term evolution of compact
binaries (cataclysmic variables and low-mass X-ray binaries).

First, we show by means of simple homology considerations that if the 
energy outflow through the surface layers of a low-mass {\rev main 
sequence} star is blocked over a fraction $s_\eff < 1$ of its surface 
(e.g. as a consequence of anisotropic irradiation) it will inflate 
only modestly, by a factor $\sim (1-s_\eff)^{-0.1}$. The maximum 
contribution to mass transfer of the thermal relaxation of the donor 
star is $s_\eff$ times what one obtains for isotropic 
($s_\eff = 1$) irradiation. The duration of this irradiation-enhanced 
mass transfer is of the order of $0.1|\ln(1-s_\eff)|$ times the thermal 
time scale of the convective envelope. Numerical computations involving 
full {\rev 1D} stellar models confirm these results. 

Second, we present a simple analytic one-zone model 
for computing the blocking effect by irradiation which gives results 
in acceptable quantitative agreement with detailed numerical
computations.  

Third, we show in a detailed stability analysis that if mass transfer 
is not strongly enhanced by consequential angular momentum losses, 
cataclysmic variables are stable against irradiation-induced runaway 
mass transfer if the mass of the main sequence donor is 
$M \la 0.7 \msol$. If $M \ga 0.7 \msol$ systems may be unstable, 
subject to the efficiency of irradiation. Low-mass X-ray binaries, 
despite providing much higher irradiating fluxes, are even less 
susceptible to this instability. 

If a binary is unstable, mass transfer must evolve through a limit
cycle in which phases of irradiation-induced high mass transfer 
alternate with phases of small (or no) mass transfer. At the peak rate
mass transfer proceeds on $s_\eff$ times the thermal time scale rate of
the convective envelope. A necessary condition for the cycles to be
maintained is that this time scale has to be much shorter ($\la 0.05$)
than the time scale on which mass transfer is driven.

\end{abstract} 

\section{Introduction} 

It is only a few years since the realization that irradiation of the
donor star of a semi-detached binary by accretion luminosity generated in
the vicinity of a compact accretor can have far-reaching consequences for 
the long-term evolution of such binary systems. Podsiadlowski (1991),
treating irradiation in spherical symmetry, first addressed this problem
in the context of the evolution of low-mass X-ray binaries (LMXBs). Later,
{\rev Gontikakis and Hameury (1993), Hameury et al. (1993), and}
Ritter (1994) pointed out that spherically symmetric irradiation is
probably not an adequate approximation for the case in
point. Preliminary calculations {\rev by Ritter (1994)} showed that the
reaction of low-mass stars to anisotropic irradiation is qualitatively 
(and quantitatively) different from that to isotropic irradiation. 
Subsequently, King et al. (1995, 1996, 1997), hereafter KFKR95, KFKR96 
and KFKR97, Ritter, Zhang and Kolb (1995, 1996), Ritter, Zhang and 
Hameury (1996), Hameury and Ritter (1997), hereafter HR97, and 
McCormick and Frank (1998), hereafter MF98, have dealt with the case of 
anisotropic irradiation in various ways and detail. In most of the 
above-quoted papers reference was made to a paper by the present authors 
in which a detailed and systematic treatment of the basic properties of 
anisotropic irradiation of low-mass stars was to be found. The 
preparation of this material has been much delayed. The main purpose of 
the present paper is to close this gap in the recent literature, and to 
provide the base for further research.  

For this purpose we shall discuss in section~2 in some detail the basic 
concepts of computing the mass transfer rate in a semi-detached binary 
with and without irradiation, and present observational evidence and 
theoretical arguments in support of the notion of anisotropic 
irradiation. In section~3 we shall present a simple analytic 
calculation which shows that the reaction of a low-mass star to 
anisotropic irradiation is qualitatively and quantitatively different 
from what one obtains in the isotropic case. We show also that the 
simple analytic results are fully supported by corresponding 
calculations of full {\rev 1D} stellar models {\rev (hereafter simply 
referred to as full stellar models)}. In section~4 we study 
the stability of mass transfer in binaries in which the donor star is 
exposed to irradiation generated by the accreting companion. In  
section~5 we shall present a more detailed semi-analytic model than 
the one in Sect.~3 for the reaction of a low-mass star to anisotropic 
irradiation in the limit of small fluxes. The implications of the 
instability against irradiation-induced runaway mass transfer for 
the long-term evolution of CVs and LMXBs is discussed in section~6. 
In section~7 we shall present and discuss results of numerical 
computations of the secular evolution of CVs {\rev subject to} the 
irradiation instability. A summary of our main results and our main 
conclusions are given in the final section~8.

\section{Computing the mass transfer rate}

In the context of this paper, compact binaries are either cataclysmic 
variables (CVs) or LMXBs, i.e. systems consisting of a low-mass star 
(the secondary) with a mass $M_s \la 1\msol$ which fills its critical 
Roche lobe and transfers matter to a compact companion (the primary), 
either a white dwarf (in CVs), or a neutron star or a black hole (in 
LMXBs). The secular evolution of such systems is a consequence of mass 
loss from the secondary which, in turn, is driven either by the 
secondary's nuclear evolution or by loss of orbital angular momentum 
and possibly other mechanisms such as irradiation on which this paper 
focuses. In the standard picture of the secular evolution see (e.g. 
King 1988; Kolb and Ritter 1992, hereafter KR92; Ritter 1996) the nature 
of the compact star is of no importance, i.e. the star is considered as 
a point mass (of mass $M_c$). The nature of the compact component is, 
however, of importance for accretion phenomena occurring in 
such systems, e.g. dwarf nova and classical nova outbursts in CVs and 
X-ray bursts in LMXBs, and when illumination of the secondary by 
radiation generated through accretion is considered, as we shall do in 
the following. Thus, calculating the standard secular evolution of such 
a binary reduces to calculating the evolution of a low-mass star under 
mass loss, where the mass loss rate derives from the boundary 
conditions imposed by the fact that the star is in a binary. 
In the simplest case one obtains the mass loss rate $\dot{M}_s$ 
by requiring that the radius of the secondary $R_s$ is always 
exactly equal to its critical Roche radius $R_R$. By decomposing 
the temporal change of $R_s$ and $R_R$ as (see e.g. Ritter 1988, 1996)
\be
\dt = \zeta_S \dmm + \pt_{\rm th} + \pt_\nuc
\label{eq1}
\ee
and
\bea
\frac{d\ln R_R}{dt} &=& \zeta_R \dmm + 
\rund{\frac{\p\ln R_R}{\p t}}_{\dm = 0}    \nonumber\\
&=& \zeta_R \dmm + 2 \rund{\frac{\p\ln J}{\p t}}_{\dm = 0}
\label{eq2}
\eea
one obtains
\bea
&&\dm = \frac{M_s}{\zeta_S - \zeta_R}\times    \nonumber\\ 
&& \eck{2 \rund{\frac{\p\ln J}{\p t}}_{\dm = 0} - \pt_{\rm th} - \pt_\nuc} \; .
\label{eq3}
\eea
Here
\be
\zeta_S = \rund{\frac{\p\ln R_s}{\p\ln M_s}}_S
\label{eq4}
\ee
is the adiabatic mass radius exponent of the secondary, and 
\be
\zeta_R = \rund{\frac{\p\ln R_R}{\p\ln M_s}}_*
\label{eq5}
\ee
the mass radius exponent of the critical Roche radius, where the 
subscript * indicates that for evaluating this quantity one has 
to specify how mass and angular momentum are redistributed in the 
system. $J$ is the orbital angular momentum, $(\p\ln R_s/\p t)_{\rm th}$ 
the rate of change of $R_s$ due to thermal relaxation and 
$(\p\ln R_s/\p t)_\nuc$ the one due to nuclear evolution. The virtue 
of (\ref{eq3}) is that it shows immediately how secular evolution works: 
If the binary is dynamically stable against mass transfer; 
i.e. if $\zeta_S-\zeta_R > 0$, then mass transfer must be driven 
by some mechanism. This can either be the secondary's expansion due 
to nuclear evolution or the shrinkage of the orbit due to loss of 
orbital angular momentum. In the standard evolution {\rev of low-mass
binaries}, mass transfer is usually not driven by thermal relaxation. 
However, as we shall see below, this is not necessarily true if 
irradiation of the secondary is taken into account.

Attempts to bring the observed population of millisecond pulsars in 
line with the death rate of their presumed progenitors, i.e. the 
LMXBs, have resulted in the speculation that the secular evolution of 
LMXBs might be drastically different from, and their lifetime much 
shorter than that of CVs (e.g. Kulkarni and Narayan 1988). A 
possible reason for this is seen in the fact that the donor in a 
LMXB is exposed to a high flux of hard X-ray radiation emitted by the 
accreting compact star. In fact, Podsiadlowski (1991) has shown that 
irradiating  a low-mass main-sequence star ($M \la 0.8 \msol$) 
spherically symmetrically with a flux $F_\irr \ga 10^{11}-10^{12}$erg 
cm$^{-2}$s$^{-1}$ results in an expansion of the star on a thermal 
time scale and in its gradual transformation into a fully radiative 
star. It is this expansion which can drive mass transfer on a thermal 
time scale and this was thought to shorten the lifetime of a LMXB. On 
the formal level the effect of irradiation is taken into account by 
replacing (\ref{eq1}) by
\bea
&&\dt = \zeta_S \dmm     \nonumber\\
&&+ \pt_\ml + \pt_\nuc + \pt_\irr \quad.            
\label{eq6}
\eea
Here, $(\p\ln R_s/dt)_\ml$ is the thermal relaxation term arising from 
mass loss alone and $(\p\ln R_s/\p t)_\irr$ the thermal relaxation term 
caused by irradiation (at constant mass). The latter term arises 
because irradiating the donor star means that its surface boundary 
conditions are changed and that it tries to adjust to them on a thermal 
time scale. With (\ref{eq6}) instead of (\ref{eq1}) we have now for the 
mass transfer rate 
\bea
\dm &=& \frac{M_s}{\zeta_S-\zeta_R} 
\Big\lbrack 2\frac{\p\ln J}{\p t} - \pt_\ml \nonumber\\ 
&-& \pt_\nuc - \pt_\irr\Big\rbrack \quad.  \label{eq7}
\eea
From (\ref{eq7}) it is clearly seen that if $(\p\ln R_s/\p t)_\irr >0$, 
irradiation amplifies mass transfer. However, there are limits to 
what irradiation can do because the time scale and the amplitude of 
the effect are limited. The secondary's expansion due to irradiation 
saturates at the latest when it has become fully radiative, where 
it is larger by $\delta \ln R_s$ than a star of the same mass in 
thermal equilibrium without irradiation. {\rev As we shall show below} 
the time scale {\rev $\tau$} on which the irradiated star initially 
expands is of the order of {\rev or less than} the thermal 
time scale of the convective envelope $\tau_{\rm ce}$. As a result, we 
can expect a maximum contribution from irradiation to $\dm$ at 
the onset of irradiation (i.e. of mass transfer) which is of the 
order 
\be
(-\delta \dm)_\irr = \frac{M_s}{\zeta_S - \zeta_R} \pt_\irr 
\la \frac{M_s}{\zeta_S - \zeta_R} \frac{\delta \ln R_s}{\tau} 
\quad.\label{eq8}
\ee
\begin{table*}
\caption[]{Detached close binaries showing a reflection effect}
\begin{tabular}{|l|c|l|c|l|l|}\hline
&associated&&&amplitude of&\\
Object& planetary& $P_{\rm orb}(d)$&$T_{\eff,1}(10^3K)$&~~reflection&Ref.\\
&nebula&&&~~~~effect&\\ \hline
BE UMa&+&2.2912&$105\pm 1$& $\Delta m_{\rm pg}=0.9$&1.2\\
VW Pyx& K1--2&0.6701&$85\pm 6$&$\Delta m=1.5$&3,4\\
V664 Cas&HFG~1&0.5817&&$\Delta B = 1.1$&5,6\\
V477 Lyr& Abell 46& 0.4717&$60\pm 10$&$\Delta V = 0.6$&7\\
UU Sge& Abell 63& 0.4651&$117.5\pm 12.5$& $\Delta V = 0.4$&8,9\\
KV Vel& DS 1& 0.3571&$77\pm 3$&$\Delta V = 0.56$& 10,11\\
TW Crv&---&0.3276&&$\Delta V = 0.85$& 12\\
AA Dor& ---& 0.2615& $40\pm 3$& $\Delta V = 0.05$& 11,13,14\\
MS Peg&---&0.1737&$22.2\pm 0.1$&$\Delta V = 0.1$&15\\
NN Ser&---& 0.1301&$55\pm 8$&$\Delta V = 0.4$&16,17\\
HW Vir&---&0.1167&$33.0\pm 0.8$&$\Delta V = 0.2$&11,18,19\\
MT Ser& Abell 41& 0.1132& $80\pm 20$& $\Delta B = 0.15$& 20,21\\
NY Vir&---&0.1010&$33\pm 3$&$\Delta V = 0.2$&22\\ \hline
\multicolumn{6}{l}{\underbar{References:}
1: Ferguson et al. (1999); 2: Wood et al. (1995); 3: Kohoutek \& Schnur}\\
\multicolumn{6}{l}{ (1982); 4: Bond \& Grauer (1987); 5: Grauer et al. (1987); 
6: Acker \& Stenholm}\\
\multicolumn{6}{l}{ (1990); 7: Pollacco \& Bell (1994); 8: Pollacco \& Bell 
(1993); 9: Bell et al. (1994);}\\
\multicolumn{6}{l}{ 10: Landolt \& Drilling (1986); 11: Hilditch et al. (1996); 
12: Chen et al. (1995);}\\
\multicolumn{6}{l}{ 13: Kilkenny et al. (1979); 14: Kudritzki et al. (1982); 
15: Schmidt et al. (1994);}\\
\multicolumn{6}{l}{ 16: Haefner (1989); 17: Catal\'an et al. (1994); 
18: Wlodarczyk \& Olszewski (1994);}\\
\multicolumn{6}{l}{ 19: Wood \& Saffer (1999); 20: Grauer \& Bond (1983); 
21: Grauer (1985);}\\
\multicolumn{6}{l}{ 22: Kilkenny et al. (1998). }\\
\end{tabular}
\end{table*}
However, with ongoing irradiation not only does $(\p\ln R_s/\p t)_\irr$ 
decrease roughly exponentially on the time scale $\tau$, but 
other effects also tend to decrease $-\dm$. With the star's 
transformation from a mainly convective to a fully radiative 
structure, $\zeta_S$ increases from $-1/3 \la \zeta_S < 0$ to a large 
positive value. Thus $\zeta_S-\zeta_R$ increases, with the result that 
$-\dm$ decreases. In addition, even the driving angular 
momentum loss is affected: for both braking mechanisms discussed in 
the literature, i.e. magnetic braking (e.g. Verbunt and Zwaan 1981; 
Mestel and Spruit 1987) and gravitational radiation (e.g. Kraft et 
al. 1962), $(\p\ln \vert \dot J\vert/\p\ln R_s) < 0$. Thus 
$\vert\dot J\vert$ decreases if the secondary inflates. Worse, 
magnetic braking which is thought to be coupled to the presence of a 
convective envelope might cease altogether once the star has become 
fully radiative. In this case the system might no longer be able to 
sustain the mass transfer necessary to keep the secondary in its 
fully radiative state. As a consequence, it is therefore possible 
that such systems evolve through a limit cycle in which short phases 
(of duration $\tau$) with irradiation-enhanced mass transfer 
alternate with long detached phases during which the oversized 
secondary shrinks as it approaches thermal equilibrium, again on the  
thermal time scale, whereas the system contracts on the much longer  
time scale of angular momentum loss. 

A number of theoretical studies along these lines, under the 
assumption of spherically symmetric irradiation, have been performed 
(e.g. by Podsiadlowski 1991; Harpaz and Rappaport 1991; Frank et al. 
1992; D'Antona and Ergma 1993; Hameury et al. 1993; Vilhu et al. 
1994), all more or less confirming the behaviour outlined above, 
including cyclic evolution (Hameury et al. 1993; Vilhu et al. 1994). 
It should be stressed once more that the validity of these studies, 
in which the effect of irradiation is maximized, rests on the validity 
of the assumption that spherically symmetric irradiation is an 
adequate approximation. In fact, Gontikakis and Hameury (1993) and 
Hameury et al. (1993) have examined whether the spherically symmetric 
approximation is adequate and find that it is not. Worse, when taking 
into account the anisotropy of the irradiation they find {\rev that} 
the long-term evolution differs significantly from the spherically 
symmetric case. Now, anisotropic irradiation is a rather difficult 
3-dimensional problem involving hydrostatic disequilibrium to some 
extent, and with it circulations which can transport heat from the hot 
to the cool side of the star. Because of this one might dismiss simple 
theoretical arguments such as those given by Gontikakis and Hameury 
(1993) and Hameury et al. (1993). There is, however, a much more 
compelling argument supporting the case of anisotropic irradiation. 
This derives from the observations of a number of close but detached 
binary systems in which a low-mass companion, exposed to an intense 
radiation field emerging from its hot (degenerate) companion, shows a 
bright illuminated and an essentially undisturbed cool hemisphere.
A compilation of the systems in question is given in Table~1, where 
we list the object's name, its association with a known planetary 
nebula, its orbital period (in days), the effective temperature of 
the irradiating white dwarf, the amplitude (in magnitudes) of the 
\lq\lq reflection effect\rq\rq, and relevant references. These 
systems all show a pronounced \lq\lq reflection effect\rq\rq in their 
light curves which, in turn, is explained by the anisotropic 
temperature distribution on the irradiated companion. Among them are 
7 binary central stars of planetary nebulae. The systems listed in 
Table~1 demonstrate that a cool star, exposed to strong irradiation 
from a hot companion, can live without problems with a hot and a cool 
hemisphere. Moreover equal effective temperature over the whole surface 
is not established, at least not over atime scale $\sim 10^4$yr 
associated with the age of central stars of planetary nebulae, despite 
the fact that the irradiated star need not even rotate nearly 
synchronously. Obviously, heat transport from the hot side to the 
cool side by means of circulations is sufficiently ineffective 
that the large difference in effective temperatures can be 
maintained over long times. Thus the case of anisotropic irradiation 
has to be taken seriously and deserves a more detailed study. This is 
the objective of this paper.

\section{Reaction of a low-mass star upon reducing its effective surface}

We start our examination of anisotropic irradiation by recalling one 
of the main results obtained in the studies of spherically symmetric 
irradiation (e.g. Podsiadlowski 1991; D'Antona and Ergma 1993), 
namely that the main effect of irradiating a low-mass main-sequence 
star is that the star cannot lose energy as effectively (or at all) 
through the irradiated parts of its surface. If irradiation is 
spherically symmetric, the star has no choice but to store the 
blocked energy in gravitational and internal energy with the well-known 
result that it swells. In the case of anisotropic irradiation, such 
as one-sided irradiation from an accreting companion, the situation 
is qualitatively different: in addition to storing the blocked 
luminosity in internal and gravitational energy, the star can also 
divert its energy flow to and lose energy from the unirradiated parts 
of its surface. This is easily possible in the adiabatic convection 
zone. Because in this zone the energy flow is almost fully decoupled 
from the mechanical and thermal structure (the flow being proportional 
to $(\nabla-\nabla_a)^{3/2}$ with $\nabla-\nabla_a \ll 1$ rather than 
to $\nabla$ (where $\nabla = \p\ln T/\p \ln P$ is the actual 
temperature gradient and $\nabla_a = (\p\ln T/\p\ln P)_a$ the adiabatic 
temperature gradient, and $T$ and $P$ are respectively the 
temperature and pressure) the mechanical and thermal structure of the 
star can still be considered to be spherically symmetric despite the 
fact that the energy flow might be highly anisotropic. The only 
thing which changes is the surface boundary condition which replaces 
the Stefan-Boltzmann law in the unirradiated case. 

In the following we shall make a simple model for studying the 
situation described above. In this model we assume that over a 
fraction $s_\eff$ of the stellar surface the energy outflow is 
totally blocked (because of irradiation, or e.g. star spots, see 
Spruit and Ritter 1983) and that the remaining fraction of the 
surface $(1-s_\eff)$ radiates with an effective temperature 
$T_\eff$. The surface luminosity of the star $L$ can therefore 
be written as 
\be
L(s_\eff) = 4\pi R^2 (1-s_\eff) \sigma
T^4_\eff\; , \quad 0 \le s_\eff < 1 \quad,\label{eq9}
\ee
where $R$ is the stellar radius. It is clear that in a more realistic 
model $s_\eff$ itself must depend on the irradiating flux $F_\irr$. 
In section~4 we shall discuss results of numerical calculations and 
a model with which we can evaluate $s_\eff(F_\irr)$. For the moment 
we note only that $s_\eff \to 0$ as $F_\irr \to 0$ and that in the 
limit of high $F_\irr$, $s_\eff$ approaches the surface fraction of 
the star which sees the irradiation source. 

The purpose of this simple modelling is, on the one hand, to 
provide estimates for the magnitude of and the time scale associated 
with the thermal relaxation process enforced by anisotropic 
irradiation, and, on the other hand, to show that the effects of 
anisotropic irradiation are not only quantitatively but also 
qualitatively different from those obtained in the spherically 
symmetric case.

The internal structure of a low-mass star with a deep outer convection 
zone can be described by the simple analytical model by Kippenhahn and 
Weigert (1994) for stars on or near the Hayashi line. This model is 
particularly applicable in our case, since we are only interested in 
the differential behaviour. Assuming a power law approximation for the 
{\rev frequency independent, i.e. grey} photospheric opacity 
$\kappa_{\rm ph}$ on the unirradiated surface of the form 
\be
\kappa_{\rm ph} = \mbox{const.} \; P^a T^b \quad,\label{eq10}
\ee
the Eddington approximation yields the photospheric solution 
\be
\log T_\eff = - \frac{a+1}{b}\log P_{\rm ph} + \frac{1}{b}\log M -
\frac{2}{b}\log R + \mbox{const.} ,\label{eq11}
\ee
where $P_{\rm ph} = P(\tau=2/3)$ is the photospheric pressure at 
an optical depth $\tau = 2/3$. The interior solution can be 
approximated by a polytrope with index $n=3/2$ and yields 
\be
\log T = \frac{2}{5} \; \log P + \frac{2}{5} \eck{\frac{3}{2} \; \log R
- \frac{1}{2} \; \log M + \mbox{const.}} .\label{eq12}
\ee
Taking (\ref{eq12}) at the photospheric point $(P=P_{\rm ph}, \; T =
T_\eff)$, i.e. equating (\ref{eq11}) and (\ref{eq12}), together with
(\ref{eq9}) and (\ref{eq10}) yields the luminosity $L$ lost by the star 
with radius $R$ over the surface area $4\pi R^2 (1-s_\eff)$: 
\be
L(s_\eff) = L_0 (1-s_\eff) \, \rund{\frac{R}{R_0}}^{(22a+4b+6)/
(5a+2b+5)}\quad . \label{eq13}
\ee
Here $L_0$ and $R_0$ are respectively the luminosity and the radius 
of the unirradiated star in thermal equilibrium. In general, the 
irradiated star will not be in thermal equilibrium, i.e. $L(R,s_\eff)$ 
does not equal the nuclear luminosity $L_\nuc$. For the disturbed star, 
the latter can be estimated by using homology relations. If we write 
the rate of nuclear energy generation $\varepsilon_\nuc$ in the form 
appropriate for hydrogen burning via the pp-chain, i.e.
\be
\varepsilon_\nuc = \mbox{const.} \; \varrho T^\nu \quad,\label{eq14}
\ee
where $\varrho$ is the density, we obtain (see e.g. Kippenhahn and 
Weigert 1994)
\be
L_\nuc (R) = L_0 \, \rund{\frac{R}{R_0}}^{-(\nu+3)}\quad.\label{eq15}
\ee
From (\ref{eq13}) and (\ref{eq15}) we obtain the gravitational luminosity  
\be
L_g (R,s_\eff) = L(R,s_\eff) - L_\nuc(R) \quad.\label{eq16}
\ee
$L_g(R,s_\eff)=0$ defines the thermal equilibrium values of the 
irradiated star. These can be expressed in terms of the thermal 
equilibrium values of the unirradiated star as follows:  
\bea
\label{eq17}
&&R_e(s_\eff) = R_0 (1-s_\eff)^r    \nonumber\\
\mbox{with}&&\\
&&r = - {5a+2b+5 \over (\nu+3)(5a+2b+5)+22a+4b+6}\quad,   \nonumber
\eea
\bea
\label{eq18}
&&L_e (s_\eff) = L_0 (1-s_\eff)^\ell   \nonumber\\
\mbox{with}&&\\
&&\ell = -(\nu+3)r \quad,   \nonumber
\eea
and
\bea
\label{eq19}
&&T_\eff(s_\eff) = T_{\eff,0}(1-s_\eff)^t    \nonumber\\
\mbox{with}&&\\
&&t = \frac{\ell-2r-1}{4} \quad.   \nonumber
\eea
Since the dominant opacity source in the photosphere of cool stars is 
$H^-$ bound-free absorption, appropriate values for $a$ and $b$ in 
(\ref{eq10}) are $a\approx 0.5$ and $b\approx4-5$. Nuclear energy 
generation in low-mass stars occurs mainly via the pp-I chain and the 
appropriate value of $\nu$ in (\ref{eq14}) is $\nu\approx 3-5$. As a 
result, we find that 
$$ r\approx -0.1\quad \nonumber $$
$$ \ell\approx 0.75 \quad, \nonumber $$
and 
$$ t\approx -0.003 \quad. \nonumber $$
This means that the effective temperature on the unirradiated part of 
the surface hardly changes, reflecting a well-known property of stars 
on or near the Hayashi line. Furthermore, since $r\approx -0.1$, the 
response of the stellar radius to anisotropic irradiation $(s_\eff<1)$ 
is much weaker than if isotropic irradiation is assumed. Specifically, 
if $s_\eff \la 0.5$, as is the case for one-sided irradiation by an 
accreting companion, the equilibrium radius with irradiation is larger 
than the one without irradiation by at most $\sim 7$\%. On the other 
hand, the total luminosity $L_e$ is significantly reduced: because of 
the rather large value of $\nu+3 \approx 6-8$ already a slight 
expansion of the star leads to a marked reduction of its nuclear 
luminosity. 

Clearly, our model is not applicable if $s_\eff\to 1$. This is 
because on the formal level $L\to 0$ if $s_\eff\to 1$ (see Eq. 
\ref{eq9}) and the values of $R$ and $T_\eff$ in thermal equilibrium 
with $r<0$ and $t<0$ (Eqs. \ref{eq17} and \ref{eq19}) diverge. There 
is also a physical reason why this model does not apply in this case: 
$s_\eff=1$ corresponds to strong, spherically symmetric irradiation 
where the star in thermal equilibrium is fully radiative (e.g. 
Podsiadlowski 1991), whereas our model applies only to the extent that 
the star in question retains a deep outer convective envelope, even 
when irradiated. 

\begin{figure}[t] 
\plotone{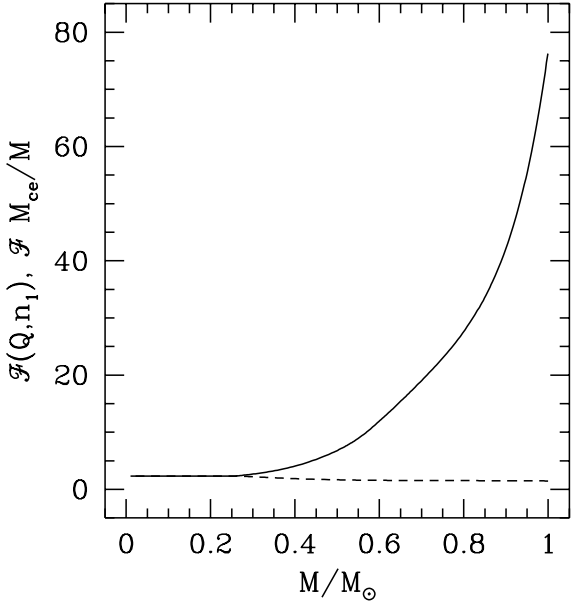}
\caption[]{The function ${\cal F}(Q,n_1)$ defined in Eq. (\ref{eq20})
(full line) and ${\cal F}(Q,n_1) M_{\rm ce}/M$ (dashed line) for
age-zero main sequence models in the bipolytrope approximation as a
function of mass.}
\end{figure}

We can now estimate the maximum contribution to mass transfer arising 
from thermal relaxation due to irradiation and the associated time 
scale. Thermal relaxation is maximal right at the onset of 
irradiation. At that time the gravitational luminosity of the star is 
$L_g(R_0,s_\eff) = -s_\eff L_0$. Using now the bipolytrope model in 
the formulation of KR92, we can write for the thermal relaxation term 
due to irradiation 
\bea
\label{eq20}
\left.\pto_\irr \right\vert_{R=R_0} &=& 
-\frac{(L_g)_\irr R_0}{GM^2} \; {\cal F}(Q,n_1) \nonumber\\
&=& \frac{s_\eff}{\tau_{\rm_{KH}}} \; {\cal F}(Q,n_1) = \frac{s_\eff}
{\tau_{\rm ce}} \quad, 
\eea
where {\rev $\tau_{\rm KH} = GM^2/R_0L_0$ is the Kelvin-Helmholtz time
of the unirradiated star in thermal equilibrium, 
$\tau_{\rm ce} = \tau_{\rm KH}/{\cal F}$ the
thermal time scale of the convective envelope, and} the quantity 
${\cal F} (Q,n_1)$ defines a dimensionless number which depends only 
on the relative size $Q = R_\core/R_0$ of and on 
the polytropic index $n_1$ in the radiative core. An explicit 
expression for ${\cal F} (Q,n_1)$ is given in KFKR96). In particular, 
for a single polytrope $n=3/2$, i.e.\ a fully convective star, where 
$Q=0$, one has ${\cal F} = 7/3$. Furthermore, we show 
${\cal F}(Q,n_1)$ as a function of mass for zero age main sequence 
stars as a full line in Fig.~1. We note that ${\cal F}$ scales roughly 
as the inverse of the relative mass {\rev $M_{\rm ce}/M$} of the 
convective envelope: the dashed line in Fig.~1 shows that 
${\cal F} \cdot M_{\rm ce}/M \approx \mbox{const.}$ within better than 
a factor of two. Thus for the purpose of an estimate we can rewrite 
(\ref{eq20}) as 
\be
\left.\pto_\irr\right\vert_{R=R_0} \approx \frac{7}{3} \; 
\frac{s_\eff}{\tau_{\rm KH}} \;\rund{\frac{M_{\rm ce}}{M}}^{-1} \quad, 
\label{eq21}
\ee
Since the radius in thermal equilibrium with irradiation is larger by
\be
\delta \ln R = \ln\, \frac{R_e(s_\eff)}{R_0} = 
r\ln (1-s_\eff) \quad,
\label{eq22}
\ee
the time scale for thermal relaxation becomes 
\bea
\label{eq23}
\tau &\approx& \delta \ln R \rund{{\p \ln R \over \p t}}_\irr^{-1} 
= \; r \; \frac{\ln (1 - s_\eff)}{s_\eff} \; \tau_{\rm ce} \nonumber\\
&\ga& r \; \frac{\ln (1 - s_\eff)}{s_\eff} \; \frac{7}{3} \; \tau_{\rm KH} \;
\frac{M_{\rm ce}}{M} \quad.
\eea
As can be seen from (\ref{eq21}) the maximal rate of expansion is 
proportional to $s_\eff$. {\rev Interestingly}, the time over which 
the new thermal equilibrium is established is much shorter than if 
$s_\eff = 1$, unless $1-s_\eff$ is a small number and our model does 
not apply anyway. 

{\rev Either (\ref{eq21}) or (\ref{eq23}) are to be inserted in
(\ref{eq8}) to get an estimate for the contribution of irradiation to
mass transfer. Using (\ref{eq21}) in (\ref{eq8}) yields the peak
contribution and (\ref{eq23}) in (\ref{eq8}) a time average.}

In order to check the validity of our simple analytical model we have 
also performed numerical computations of full   stellar models using 
the modified Stefan-Boltzmann law (\ref{eq9}) as one of the outer 
boundary conditions and with $s_\eff$ as a free parameter. For our 
computations we have used a modified version of Mazzitelli's (1989) 
stellar evolution code which is described in more detail in KR92, and 
have assumed a standard Pop. I chemical composition with (in the 
usual notation) $X=0.70$ and $Z=0.02$.

\begin{figure}[t] 
\plotone{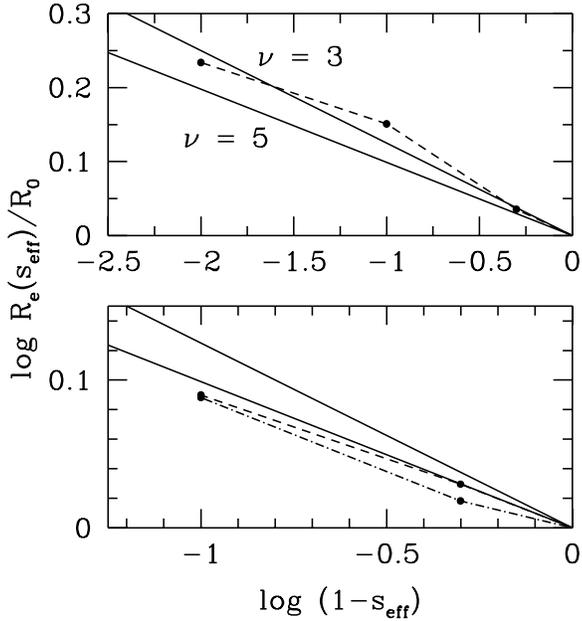}
\caption[]{Radius $R_e(s_\eff)$ of homogeneous stellar models in 
thermal equilibrium as a function of the fraction $s_\eff$ of the 
stellar surface over which energy loss is assumed to be blocked. 
{\rev The} full lines in both panels show the prediction of the 
homology model (Eq. \ref{eq17}) with the parameters $a=0.5$, $b=4$, 
and $\nu$ as indicated. The dashed line in the upper panel is for 
full stellar models with a mass of $0.4\msol$, the dashed and dotted 
lines in the lower panel for full stellar models with a mass of 
respectively $0.2\msol$ and $0.8\msol$.}
\end{figure}

One of the basic predictions of our simple model is that (see 
Eq. \ref{eq17}) $\log (R_e(s_\eff)/R_0)$ scales linearly with $\log 
(1-s_\eff)$ and that the slope $\vert r\vert$ is small. This is 
nicely confirmed by the behaviour of full stellar models shown in 
Fig.~2. As can be seen, the prediction is valid, at least 
qualitatively, over more than two orders of magnitude of 
$(1-s_\eff)$. Furthermore the slope in Fig.~2 is indeed small, 
confirming that $\vert r\vert$ is a small number. That the slope is 
different for stars of different mass is due to the fact that the 
effective values of the parameters $a$, $b$ and $\nu$ change with 
stellar mass. 

\begin{figure}[t] 
\plotone{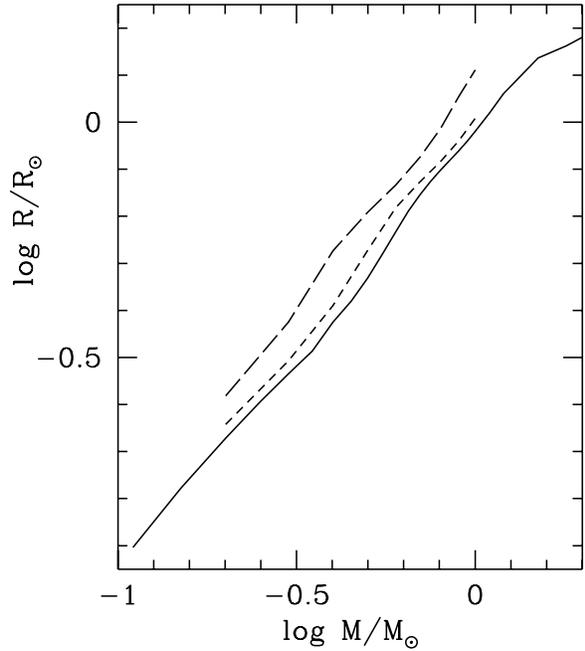}
\caption[]{Mass radius diagram of full stellar models in thermal
equilibrium forr three different values of the fraction $s_\eff$ of
the stellar surface over which energy loss is assumed to be blocked.
Full line: $s_\eff=0$ (normal main sequence); short dashed line:
$s_\eff=0.5$, and long dashed line: $s_\eff=0.9$.}
\end{figure}

In Fig.~3 we show in a mass radius diagram the thermal equilibrium 
radius $R_e$ as a function of mass for three different values of 
$s_\eff$, i.e.\ for the standard main sequence $(s_\eff = 0)$, and 
for $s_\eff = 0.5$ and $s_\eff = 0.9$. As can be seen, the mass 
radius relations for $s_\eff=0.5$ and $s_\eff = 0.9$ are shifted by 
a small amount to larger radii and run roughly \lq\lq parallel\rq\rq 
to the standard main sequence $(s_\eff = 0)$, as is predicted by our 
simple model. 

\begin{figure}[t] 
\plotone{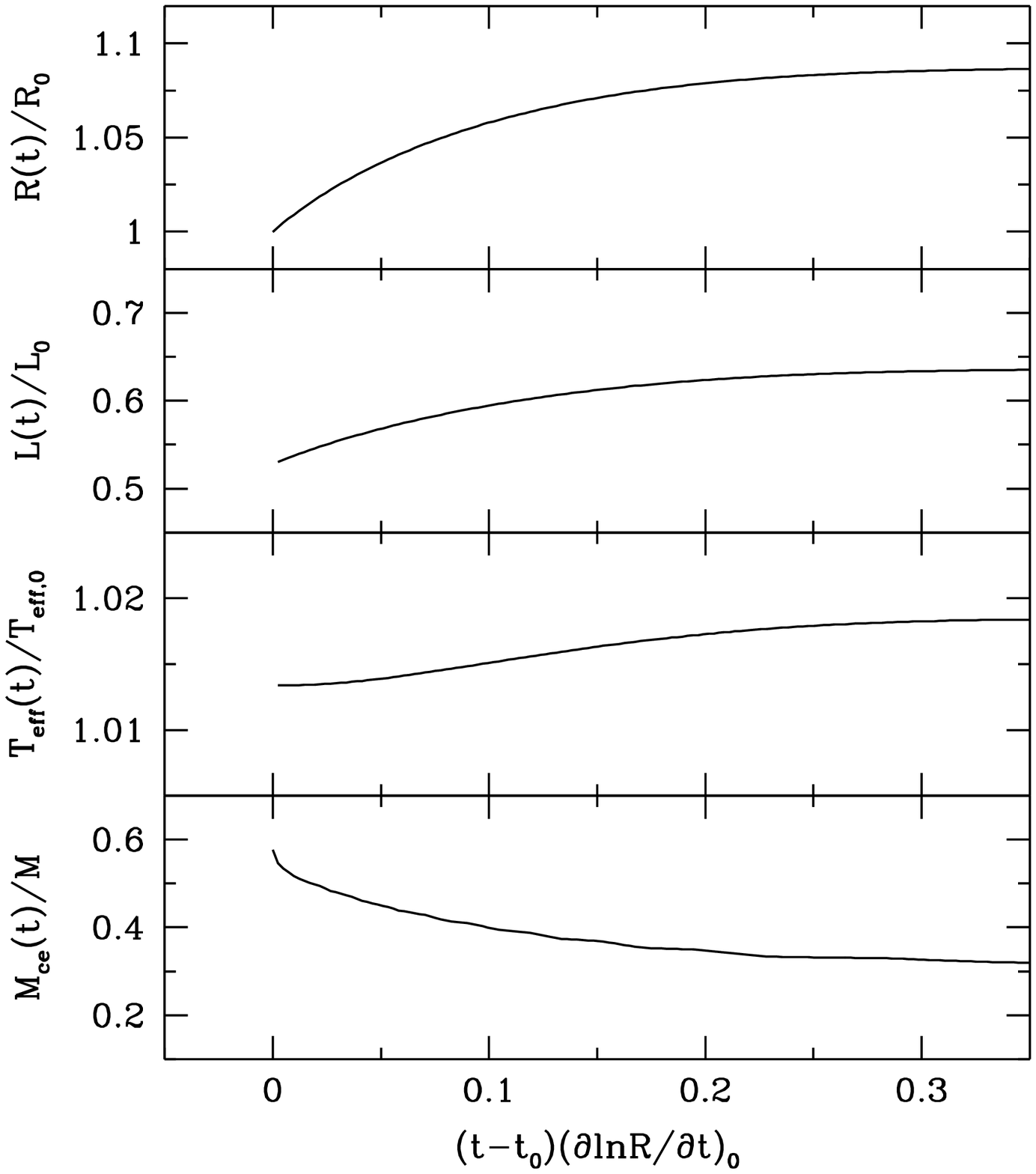}
\caption[]{Thermal relaxation of a $0.4\msol$ main sequence star
after blocking the energy loss over a fraction $s_\eff=0.5$ of its surface 
at time $t=t_0$. Top frame: radius $R$, second frame: luminosity $L$,
third frame: effective temperature $T_\eff$ of the radiating part, and 
bottom frame: the relative mass $M_{\rm ce}/M$ of the convective envelope
as a function of time. Time is measured in units of the time scale on
which the radius grows at $t=t_0$. $R_0$, $L_0$, and $T_{\rm eff,0}$ are 
respectively the values of $R$, $L$ and $T_\eff$ immediately before the 
onset of the blocking of energy outflow.}
\end{figure}

Finally, in Fig.~4 we show as an example the thermal relaxation  
{\rev with} time of an $0.4\mo$ star with $s_\eff = 0.5$. If time is 
measured in units of $(\p t/\p\ln R)\vert_{R=R_0}$, as is done in 
Fig.~4, it is seen that the relaxation process is characterized by this 
time scale and that new thermal equilibrium is reached after only 
0.2--0.3 of these time units. Again this confirms our analytical result 
(cf. Eq. \ref{eq23}), according to which the relaxation process lasts a 
few $r\ln(1-s_\eff)$ in these units. It is also seen that the effective 
temperature on the unirradiated part of the star rises only very little, 
as predicted, but that the relative mass of the convective envelope is 
reduced significantly from $\sim 0.6$ at the beginning to $\sim 0.33$ 
in the new thermal equilibrium.

\section{Stability against irradiation-induced mass transfer}

Let us now examine the situation in which a low-mass star transfers 
mass to a compact companion (of mass $M_c$ and radius $R_c$) and, in 
turn, is irradiated (directly or indirectly) by the accretion light 
source. Because irradiation can enhance mass transfer and more 
irradiation can give rise to even higher mass transfer, we must 
examine under which conditions such a situation is stable against 
irradiation-induced runaway mass transfer. 

\subsection{Arbitrary irradiation geometry}
In this subsection we wish to keep the discussion as general as 
possible. Therefore, we do not specify the irradiation geometry. 
Specific models which do just that will be presented in the next 
subsection (4.2). 

The component normal to the stellar surface of the irradiating flux 
generated by accretion can be written as 
\be
F_\irr (\vartheta,\varphi) = \frac{\eta}{4\pi}\; 
\frac{G M_c (-\dm)}{R_c \, a^2} \; 
h(\vartheta,\varphi) \quad.\label{eq24} 
\ee
Here $-\dm$ is the mass transfer rate, $a$ the orbital separation, 
$h(\vartheta,\varphi)$ a dimensionless function of the position on 
the secondary's surface (characterized by polar coordinates 
$(\vartheta,\varphi)$) which describes the irradiation geometry, 
and $\eta<1$ a dimensionless efficiency factor. $\eta = 1$ only if 
the accretion luminosity is radiated isotropically from the compact 
star and if all the energy which is radiated into the solid angle 
subtended by the donor as seen from the accretor is absorbed below 
the photosphere. In a real situation $\eta$ will be considerably 
less than unity for several reasons. The most important of these 
are: 1) the accretion luminosity will, in general, not be emitted 
isotropically, e.g. if accretion occurs via a disk which radiates 
predominantly perpendicular to the orbital plane and casts a shadow 
onto the donor, or if the accretor is strongly  magnetized and 
accretes mainly near the magnetic poles, as e.g. in AM Her systems. 
2) energy emitted in certain spectral ranges, as e.g. EUV radiation 
and soft X-rays, is unlikely to reach the photosphere of the donor 
(in the case of EUV and soft X-ray radiation because of the high 
column density of neutral hydrogen). 3) part of the incident flux 
will be scattered away before penetrating into the photosphere. 4) 
not all of the transferred mass needs to be accreted by the compact 
star. Part of it may leave the system before releasing much potential 
energy, e.g. via a wind from the outer regions of the accretion disk. 

From {\rev this} it is clear that computing a reliable value for 
$\eta$ is a formidable if not nearly impossible task. Therefore, we  
treat $\eta$ as a free parameter and our goal must be to arrive at 
conclusions which are as independent of $\eta$ as possible. 

For deriving the criterion for adiabatic stability against  
irradiation-induced mass transfer we follow Ritter (1988) (see 
also Ritter 1996). The only difference is that we use now 
Eq.~(\ref{eq6}) instead of (\ref{eq1}) (the latter corresponding to 
Eq.~(12) in Ritter's (1988) paper). Accordingly the criterion for 
adiabatic stability becomes 
\bea
\label{eq25}
\zeta_S - \zeta_R > \zeta_\irr &=& -M_s \; 
\frac{\p}{\p\dm} \; \pt              \nonumber\\
&=& -M_s \; \frac{\p}{\p L} \; \pt \; \frac{dL}{d\dm} \quad.
\eea
The dimensionless number $\zeta_\irr$ which is defined by (\ref{eq25}) 
measures how sensitively the stellar radius changes in response to 
irradiation produced by mass transfer. Ignoring the influence of 
irradiation, i.e. setting $dL/d\dm = 0$, gives $\zeta_\irr = 0$ and 
(\ref{eq25}) reduces to the usual criterion for adiabatic stability. 

We compute the derivative $(\p^2\ln R_s/\p t \p L)$ in (\ref{eq25}) 
in the framework of the bipolytrope model (e.g. KR92) from which we 
obtain (cf Eq.~20)
\be
\frac{\p}{\p L} \; \pt = - \frac{R_s}{GM^2_s} \; 
{\cal F}(Q,n_1) \quad.\label{eq26}
\ee
For calculating $dL/d\dm$ we shall make use of what we shall refer 
to as the weak irradiation assumption. Making this assumption is 
tantamount to assuming that at any point $(\vartheta,\varphi)$ 
lateral heat transport is negligible compared to radial transport. 
Lateral heat transport occurs in the form of radiative diffusion 
and advection because of non-vanishing lateral temperature gradients 
$\p T/\p\vartheta$ and $\p T/\p\varphi$, and departures from strict 
hydrostatic equilibrium. We shall show in the Appendix that the weak 
irradiation assumption can be justified in those cases we are 
interested in and that we may safely neglect lateral heat transport. 
Accordingly, energy conservation  requires that at any point 
$(\vartheta,\varphi)$ the stellar flux, i.e. the energy lost by the 
star from its interior per unit time and unit surface area is
\be
F(\vartheta,\varphi) = \sigma T^4_\irr (\vartheta,\varphi) - F_\irr
(\vartheta,\varphi) \quad,\label{eq27}
\ee
where $T_\irr$ is the effective temperature of the surface element 
in question. With (\ref{eq27}) the stellar luminosity, i.e. the 
energy loss per unit time from the interior becomes
\be
L = R^2_s \int^{2\pi}_0 \int^\pi_0 F(\vartheta,\varphi) \; \sin
\vartheta d \vartheta \, d \varphi \quad.\label{eq28}
\ee
Before working out $dL/d\dm$ in (\ref{eq25}), we shall first examine 
the reaction of the stellar surface to irradiation in rather general 
terms. For that it is convenient to introduce the dimensionless 
irradiating flux
\be
x = \frac{F_\irr}{\sigma T_0^4} = \frac{F_\irr}{F_0} \label{eq29}
\ee
and the dimensionless stellar flux
\be
G = \frac{F}{F_0} = \rund{\frac{T_\irr (x)}{T_0}}^4 - x =
G(x)\quad,\label{eq30}
\ee
where $T_0 = T_\eff (F_\irr = 0)$ is the effective temperature and 
$F_0 = F(F_\irr = 0) = \sigma T^4_0$ the stellar flux in the absence 
of irradiation. We note that $G(0)=1$ and that we expect $G(\infty)=0$, 
i.e. that for very high irradiating fluxes energy outflow from the 
stellar interior is totally blocked. Next we introduce the function 
\be
g(x) = -\frac{dG}{dx} = -\frac{dF}{dF_{\rm irr}} \quad,\label{eq31}
\ee
which has the following notable properties: First 
\be
\int^\infty_0 g(x) dx = G(0) - G(\infty) = 1\label{eq32}
\ee
Second, for positive albedos 
\be
0 < g(x) < 1 \quad \forall\, x \ge 0\quad.\label{eq33}
\ee
Third, if $F$ is a monotonically decreasing function of $x$ (and 
there is no physical reason why this should not be so), then 
\be
g^\prime(x) < 0 \quad \forall\, x \ge 0 \quad.\label{eq34}
\ee
From (\ref{eq32}) and (\ref{eq34}) we can immediately prove through 
integration of $g(x)$ by parts that
\be
\mbox{Max} \eck{xg(x)} < 1 \quad.\label{eq35}
\ee
{\rev The reason why $xg(x)$ attains a maximum can be understood as
follows: Rewriting $xg(x)$ in dimensional form (using (\ref{eq29}) 
and (\ref{eq31})) we see that 
\be
\label{eq35a}
xg(x) = - \frac{F_{\rm irr}}{F_0} \; \frac{dF}{dF_{\rm irr}} \quad.
\ee
The second factor in (\ref{eq35a}) describes the incremental blocking
of the energy loss from the interior with changing irradiating flux
$F_{\rm irr}$. The maximum of $xg(x)$ arises because $-{dF}/{dF_{\rm
irr}}$ vanishes for large $F_{\rm irr}$. This, in turn, is a
consequence of the fact that as long as the star keeps a negative
temperature gradient $dT/dr$ in its subphotospheric layer, i.e. the
superadiabatic convection zone, irradiation can not block more than
the total flux $F_0$.}

As we shall see below, the fact that $xg(x)$ has a maximum is very 
important for the stability discussion. In fact, we shall see in the 
next section 5 that for realistic situations the maximum of $xg(x)$ 
is smaller by about a factor of two than the strict upper limit given 
by (\ref{eq35}).

We can now return to Eq. (\ref{eq28}) and compute $dL/d\dm$. This 
can be written as 
\be
\label{eq36}
\frac{dL}{d\dm} = 
-R^2_s \int^{2\pi}_0\int^\pi_0 \frac{dF}{dF_\irr} \; 
\frac{dF_\irr}{d(-\dm)} \; \sin \vartheta d\vartheta d\varphi ,
\ee
where we note that the first factor in the integral {\rev is} equal 
to $-g$. With
$$
\frac{dF_\irr}{d(-\dm)} = \frac{F_\irr (\vart,\varp)}{(-\dm)}\eqno(38a)
$$
$$
= \frac{\eta}{4\pi}\frac{GM_c}{R_c a^2} \; h(\vart,\varp)\eqno(38b)
$$
\stepcounter{equation}
from (\ref{eq24}) we have {\rev (using (38a))}
\bea
\label{eq38}
&&\frac{dL}{d\dm} = \frac{1}{4\pi}\frac{L_0}{(-\dm)} 
\rund{\frac{R_s}{R_0}}^2 \times \nonumber\\
&& \int^{2\pi}_0\int^\pi_0 x(\vart,\varp) \, 
g(x(\vart,\varp)) \, \sin\vart d\vart d\varp ~,
\eea
where {\rev $L_0 = 4\pi R^2_0 \sigma T_0^4$} is the luminosity of 
the star in thermal equilibrium without irradiation. Combining now 
Eqs. (\ref{eq25}), (\ref{eq26}) and (\ref{eq38}) we can rewrite the 
stability criterion as 
\bea
\label{eq40}
&&\zeta_S - \zeta_R > \zeta_\irr =      
\frac{1}{4\pi}\frac{\tau_{M_s}}{\tau_{\rm KH}} \; 
{\cal F} (Q,n_1) \rund{\frac{R_s}{R_0}}^3 \times \nonumber\\
&&\int^{2\pi}_0\int^\pi_0 x(\vart,\varp)\, g(x(\vart,\varp)) \, 
\sin\vart d\vart d\varp
\eea
or
\bea
\label{eq41}
&&\Lambda \equiv 2(\zeta_S-\zeta_R) \; \frac{\tau_{\rm KH}}{\tau_{M_s}} \, 
{\cal F}^{-1} (Q,n_1) \rund{\frac{R_s}{R_0}}^{-3}       \nonumber\\
&&> \frac{1}{2\pi} \int^{2\pi}_0\int^\pi_0 x(\vart,\varp) \, 
g(x(\vart,\varp)) \, \sin\vart d\vart d\varp \equiv I \quad,
\eea
where
\be
\tau_{M_s} = \frac{M_s}{-\dm} \label{eq42}
\ee
is the mass loss time scale. Although the relations (\ref{eq40}) and 
(\ref{eq41}) do not show an explicit dependence on $\eta$ they 
nevertheless depend on it via $x$. However, because of the fact that 
$xg(x)$ has a maximum, the integral on the right-hand side of 
(\ref{eq40}) and (\ref{eq41}) must have a maximum that is smaller 
than Max$(xg(x))$. Hence we can state that systems which fulfill the 
condition
\be
\Lambda > \frac{1}{2\pi} \mbox{Max} 
\eck{\int^{2\pi}_0\int^\pi_0 x(\vart,\varp) g(x(\vart,\varp)) 
\sin\vart d\vart d \varp}\; ,\label{eq44}
\ee
which {\it is} independent of $\eta$, are definitely 
stable against \break irradiation-induced runaway mass transfer. 

The {\rev reason for} rewriting the stability criterion (\ref{eq40}) 
in the form of (\ref{eq41}) or (\ref{eq44}) is that in the latter 
conditions $\Lambda$ does not depend on irradiation but only on the 
internal structure of the donor star (via $\zeta_S$, $\tau_{\rm KH}$, 
${\cal F}(Q,n_1)$, $R_0$) and on the secular evolution model (via 
$\tau_{M_s}$, $\zeta_R$ and $R_s$). On the other hand, all the 
information about the irradiation model is contained in the 
expression on the right-hand side.

{\rev For discussing} the stability of mass transfer in the limit of 
very small irradiating fluxes, i.e. $x\to 0$, we must use (38b) 
instead of (38a) in (\ref{eq38}). This results in the following 
stability criterion:
\bea
\label{eq45}
&&\zeta_S - \zeta_R > \zeta_\irr = \frac{\eta}{4\pi} \frac{M_c}{M_s} 
\frac{R_s}{R_c}\rund{\frac{R_s}{a}}^2 \; {\cal F}(Q,n_1)\times \nonumber\\
&& \iint g(x(\vart,\varp)) \, h(\vart,\varp) \, \sin\vart d\vart d\varp \quad.
\eea
or
\bea
\label{eq46}
\Gamma &\equiv& 2(\zeta_S-\zeta_R) \; \frac{M_s}{M_c}\frac{R_c}{R_s}
\rund{\frac{a}{R_s}}^2 {\cal F}^{-1}(Q,n_1) \nonumber\\
&>& \frac{\eta}{2\pi} \iint g(x(\vart,\varp)) \, h(\vart,\varp) \, 
\sin \vart d\vart d\varp\quad.
\eea
Because of (\ref{eq34}) we arrive at a necessary and sufficient 
condition for stability 
\be
\Gamma > \frac{\eta}{2\pi} \, g(0) \iint h(\vart,\varp)\, 
\sin\vart d\vart d\varp\quad.\label{eq47}
\ee
As above, we have separated in the conditions (\ref{eq46}) and 
(\ref{eq47}) factors which do not depend on irradiation (collected 
in $\Gamma$) from those which do (on the right-hand side). Like 
$\Lambda$, $\Gamma$ depends only on the internal structure of the 
donor star and the secular evolution model. In essence, Eqs. 
(\ref{eq46}) and (\ref{eq47}) are conditions on the value of $\eta$ 
in the sense that for a given irradiation model, i.e. given 
$h(\vart,\varp)$ and $g(x)$, $\eta$ must not exceed a certain value 
if mass transfer is to be stable. 

Furthermore, Eqs. (\ref{eq46}) and (\ref{eq47}) show in particular 
that because $g(x)$ is maximal for $x=0$ (see Eq. \ref{eq34}), i.e. 
for $F_\irr = 0$, systems with an unirradiated donor star are the 
most susceptible to irradiation. In other words, if a system is 
stable at the turn-on of mass transfer it will remain so later, 
unless secular effects diminish the value of $\Gamma$.

\subsection{Specific irradiation models} 
In the following we describe two specific irradiation models, a 
rather simple one, hereafter referred to as the constant flux or 
constant temperature model, and a more realistic one, hereafter 
referred to as the point source model.

\subsubsection{The constant flux model} 
In this model we assume a fraction $s(\approx 0.5)$ of the 
stellar surface to be irradiated by a constant average normal flux
\be
\langle F_\irr \rangle = \frac{\eta}{8\pi}\frac{GM(-\dm)}{R_c \, a^2} 
\quad. \label{eq49}
\ee
Note that the parameter $s$, introduced above, and $s_\eff$ which 
we have introduced in Sect.~3 are not the same quantity. However, 
$s$ and $s_\eff$ are related and the corresponding relation will be 
given below. 

Comparison of (\ref{eq49}) with (\ref{eq24}) shows that we may write 
$h(\vart,\varp)$ as follows:
\be
h(\vart,\varp)=
\left\{ \begin{array}{lll}
\frac{1}{2}\; ,\quad 0 \le \vart < \vart_\max\; , 
\quad 0 \le \varphi \le 2\pi \\ 0\; , \quad \vart_\max \le \vart \le \pi \; , 
\quad 0 \le \varphi \le 2\pi
\end{array}\right.\quad,\label{eq50}
\ee
where now $\vart$ is the colatitude with respect to the substellar 
point and $\varp$ the azimuth around the axis joining the two stars. 
If the star is assumed to be spherical, the colatitude $\vart_\max$ 
of the \lq\lq terminator\rq\rq of the irradiated part of the surface 
and $s$ are related via
\be
s = {1\over 2} \, (1-\cos\vart_\max)\quad.\label{eq51}
\ee
With (\ref{eq50}) and (\ref{eq51}), and $\langle x\rangle = 
\langle F_\irr\rangle/F_0$, the stability conditions (\ref{eq41}) 
and (\ref{eq46}) become respectively 
\be
\Lambda > 2s \langle x\rangle g (\langle x\rangle)\label{eq52}
\ee
and
\be
\Gamma > \eta \, s \, g(\langle x\rangle)\quad.\label{eq53}
\ee
These are the results presented earlier in Ritter, Zhang and 
Kolb (1995, 1996). 

Because of (\ref{eq50}) not only is the irradiating normal flux 
constant but also the effective temperature on the irradiated part 
(hence the name constant temperature model). (\ref{eq50}) inserted 
in (\ref{eq28}) yields the luminosity of the star
\be
L = 4\pi R^2_s \eck{s (\sigma T^4_\irr - \langle F_\irr\rangle) + 
(1-s) \sigma T^4_0}\quad.\label{eq54}
\ee
Comparing this modified Stefan-Boltzmann law with (\ref{eq9}) 
yields the relation between $s$ and $s_\eff$: 
\be
s_\eff = s \eck{1-\frac{\sigma T^4_\irr - \langle F_\irr\rangle}
{\sigma T^4_0}} = s \eck{1-G(\langle x\rangle)}\quad.\label{eq55}
\ee

\begin{figure}[t] 
\vspace*{-2.5cm}
\plotone{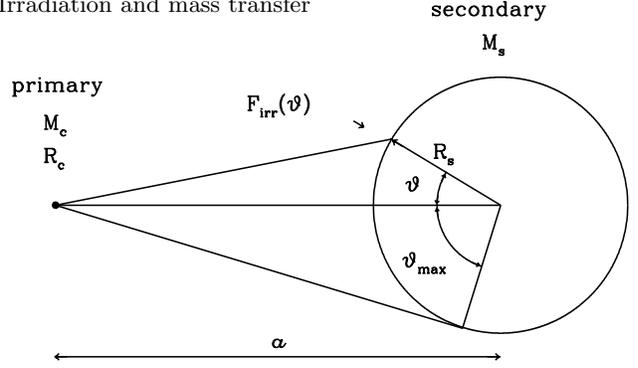}
\caption[]{Sketch of the geometry of the point source model involving
a spherical secondary irradiated by a point source at a distance $a$.
Note that $F_\irr(\vartheta)$ is the irradiating flux normal to the 
stellar surface. }
\end{figure}

\subsubsection{The point source model} 
In this model, which has already been discussed in some detail by 
KFKR96, we assume the donor star to be irradiated by a point source 
at the location of the compact star. For simplicity we assume the 
secondary to be spherical. The chosen geometry is axisymmetric with 
respect to the axis joining the two stars. Denoting again by $\vart$ 
the colatitude of a point on the surface of the irradiated star with 
respect to the substellar point (for a sketch of the geometry see 
Fig.~5), $h(\vart,\varp)$ in (\ref{eq24}) becomes
\be
h(\vart) = \frac{\cos \vart-f_s}
{(1-2f_s \cos \vart + f^2_s)^{3/2}}\quad,\label{eq56}
\ee
where
\be
f_s = {R_s \over a} = f_s\rund{\frac{M_c}{M_s}} \label{eq57}
\ee
is the secondary's radius in units of the orbital separation $a$. 
Because the secondary fills its critical Roche volume, $f_s$ is a 
function only of the mass ratio $M_c/M_s$. The terminator of the 
irradiated part of the star is at the colatitude 
\be
\vart_\max = \mbox{arc}\cos(f_s)\quad.\label{eq58}
\ee
Inserting (\ref{eq56}) in (\ref{eq41}) or (\ref{eq46}), the stability
criteria become respectively
\be
\Lambda > \int^{\vart_\max}_0 x(\vart) \, g (x(\vart)) \, 
\sin\vart d\vart = I_{\rm PS}\label{eq59}
\ee
and
\be
\Gamma > \eta \int^{\vart_\max}_0 g(x(\vart))\, h(\vart)\, 
\sin\vart d\vart\quad.\label{eq60}
\ee
With (\ref{eq56}) inserted in (\ref{eq28}), the luminosity of the star 
is
\bea
\label{eq61}
L &=& 4\pi R^2_s \sigma T^4_0 \times \nonumber\\
&&\left\{ \frac{1}{2} (1+f_s) + \frac{1}{2} \int^{\vart_\max}_0 G(x(\vart))\, 
\sin\vart \, d\vart\right\}\quad.
\eea
Comparing (\ref{eq61}) with (\ref{eq9}) we find that the effective
fraction of the stellar surface over which the energy outflow is 
blocked is
\be
s_\eff = \frac{1}{2} \left\{1-f_s-\int^{\vart_\max}_0 G(x(\vart))\, 
\sin\vart \, d\vart\right\}\quad.\label{eq62}
\ee

\section{The reaction of the subphotospheric layers to irradiation} 

From the stability analysis we have carried out in the previous 
section it is clear that we need to know more about the functions 
$G(x)$ or $g(x)$ (cf. Eqs.~30 and 31) if we wish to use the stability 
criteria in a quantitative way. So far we know only the properties 
detailed in Eqs.~(\ref{eq32})--(\ref{eq35}). These, however, are 
insufficient for our purposes. Therefore, in this section we shall 
first use a simple model to derive $g(x)$ explicitly and thereafter 
discuss results of numerical calculations.As in section~4 we shall 
adopt the weak irradiation assumption. In this approximation the 
relation between the stellar flux $F$ and the irradiating flux 
$F_\irr$ is a purely local property. 

\subsection{A one-zone model for the superadiabatic layer}
For determining $F(F_\irr)$ we need now a more detailed model of 
the stellar structure than the one we have assumed in Sect.~3. 
Whereas in Sect.~3 we have assumed with Kippenhahn and Weigert 
(1994) that the convective envelope remains adiabatic up to the 
photosphere, we shall now relax this assumption and take into 
account the existence of a thin superadiabatic convection zone 
below the photosphere where convection itself is ineffective as 
a means of energy transport and energy flows mainly via radiative 
diffusion. Sufficiently deep in the star, where convection is 
effective, i.e. adiabatic, the thermal and mechanical structure 
of the envelope remain spherically symmetric to a very good 
approximation even in the presence of anisotropic irradiation 
at the surface (cf. our discussion in Sect.~2). It is only the 
very thin superadiabatic layer (with a mass of typically $10^{-10}\msol$), 
where energy is mainly transported via radiation, which is strongly 
affected by irradiation. It is this property which allows us to make 
the weak irradiation assumption, i.e. to treat the effects of 
irradiation in a local approximation and with the following simple model.

\begin{figure}[t] 
\vspace*{-1.5cm}
\plotone{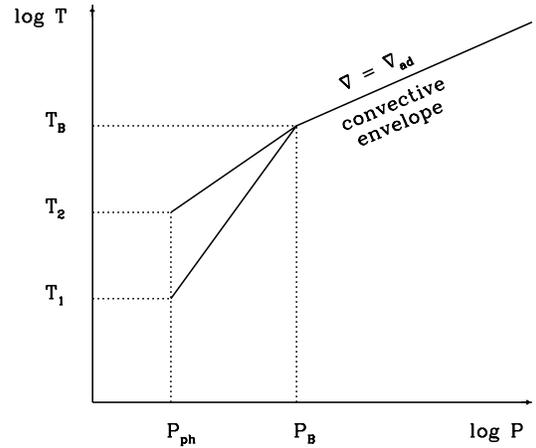}
\caption[]{Sketch of the temperature pressure stratification adopted
in the framework of the one-zone model described in Sect. 5.1.}
\end{figure}

For this we adopt for the moment the temperature-pressure 
stratification shown in Fig.~6. This means that we replace the 
``true" structure by one in which we assume convection to be 
adiabatic out to a point where the pressure is $P=P_B$ and the 
temperature $T=T_B$ (where the subscript $B$ stands for the base of 
the superadiabatic zone). The superadiabatic convection zone extends 
from the point $(P_B,T_B)$ up to the photosphere where $P=P_{\rm ph}$ 
and $T=T_\eff$. Because in this zone convection is ineffective, we make 
the simplifying assumption that energy transport is via radiation only, 
i.e.\ that $\nabla=\nabla_\rad$, where $\nabla_\rad$ is the radiative 
temperature gradient. Because on an irradiated part $T_\eff = T_\irr 
> T_0$, where $T_0$ is the effective temperature in the absence of 
irradiation, but $T_B$ is assumed to be the same irrespective of 
irradiation, we see that irradiation, by raising the effective 
temperature, reduces the temperature gradient $\nabla$ and thus the 
radiative energy loss through these layers. The superadiabatic layer 
works like a valve which is open if there is no external irradiation 
and which closes in progression with the irradiating flux. As sketched 
in Fig.~6 we assume for simplicity that $P_{\rm ph}$ does not depend on 
$F_\irr$, i.e.\ on $T_\eff$.

We can now to derive $T_\irr(F_\irr)$ by making a simple one-zone model 
for the superadiabatic layer. For this we shall furthermore assume that 
the optical depth through this layer is large enough for the 
diffusion approximation to hold. Then the radiative flux is given by 
\be
F_\rad = F = -\frac{ac}{3\kappa\varrho}
\frac{dT^4}{dr}\quad,\label{eq63}
\ee
where $\varrho$ is the density, $\kappa$ the opacity and the other 
symbols have their usual meaning. 

Now we consider the transported flux in two different superadiabatic 
layers (using subscripts 1 and 2) on an arbitrary isobar at some level
$P_{\rm ph} < P_1 = P_2 < P_B$. The radiative fluxes are then 
\be
F_i = -\frac{ac}{3\kappa_i\varrho_i} 
\rund{\frac{dT^4}{dr}}_i \; , \quad i = 1,2 \quad.\label{eq64}
\ee
Therefore
\be
\frac{\kappa_1 \varrho_1}{\kappa_2\varrho_2} = \frac{F_2}{F_1} \; 
\frac{(dT^4/dr)_1}{(dT^4/dr)_2}\quad.\label{eq65}
\ee
Now we use for the opacity a power-law approximation of the form 
(\ref{eq10}) and the ideal gas equation 
\be
P = \frac{{\cal R}}{\mu}\;\varrho T\quad,\label{eq66}
\ee
where ${\cal R}$ is the gas constant and $\mu$ the mean molecular 
weight. Since in the regions of interest the dominant species (H and 
He) are neutral, we may assume $\mu_1=\mu_2$. Thus with $P_1=P_2$ we 
have
\be
\varrho_1 T_1 = \varrho_2 T_2 \quad.\label{eq67}
\ee
Inserting now (\ref{eq10}) and (\ref{eq67}) into (\ref{eq65}) we obtain
\be
\frac{F_1}{F_2} = \frac{T^{1-b}_1(dT^4/dr)_1}{T^{1-b}_2(dT^4/dr)_2} = 
\left\{\begin{array}{cc}
\frac{(dT^{5-b}/dr)_1}{(dT^{5-b}/dr)_2} \; ,& 5-b\ne 0\\
&\\
\frac{(d\ln T/dr)_1}{(d\ln T/dr)_2}\; ,& 5-b=0
\end{array}\right.\; .\label{eq68}
\ee
Now we make the one-zone approximation by writing 
\bea
\label{eq69}
& \frac{dT^n}{dr} = \frac{T^n(P=P_B) - T^n (P=P_{\rm ph})}{\Delta r} = 
\frac{T^n_B - T^n_\eff}{\Delta r} & \nonumber\\
&&\\
& \frac{d\ln T}{dr} = \frac{\ln T(P=P_B) - \ln T(P=P_{\rm ph})}{\Delta r} =
\frac{\ln T_B - \ln T_\eff}{\Delta r} \quad.&\nonumber
\eea
If we now identify layer 1 with the unirradiated one, i.e. set 
$F_1=F_0$ and $T_{\eff,1} = T_0$, and layer 2 with an irradiated 
one, i.e.\ set $F_2 = F$ and $T_{\eff,2} = T_\irr$ we obtain by 
inserting (\ref{eq69}) into (\ref{eq68}) 
\be
\frac{F_0}{F} = \frac{1}{G} = \frac{\sigma T_0^4}{\sigma T^4_\irr - F_\irr} 
= \left\{ \begin{array}{ll}
\frac{T_B^{5-b} - T^{5-b}_0}{T^{5-b}_B - T^{5-b}_\irr} \; ,&b-5 \ne 0\\
&\\
\frac{\ln T_B - \ln T_0}{\ln T_B - \ln T_\irr} \; ,& b-5 = 0
\end{array}\right. ,\label{eq70}
\ee
where we have assumed for simplicity $\Delta r_1 = \Delta r_2$. 
Equation (\ref{eq70}) can be solved for 
$T_\irr = T_\irr (T_0,T_B,F_\irr)$, thus providing 
$G(x,T_B)$. Thus (\ref{eq70}) together with (\ref{eq28}) or special 
cases thereof (Eqs. \ref{eq54} or \ref{eq61}) can be used as an outer 
boundary condition for numerical computations. 
 
Let us now briefly consider the consequences of our above assumptions 
that $P_{\rm ph,1} = P_{\rm ph,2}$ and $\Delta r_1 = \Delta r_2$. 
Because the dominant opacity source is H$^-$ and therefore the opacity 
increases steeply with temperature, the photospheric pressure on an 
irradiated, hotter part of the star will be lower than on an 
unirradiated part (because $P_{\rm ph} \sim g/\kappa$, where $g$ is 
the surface gravity). Furthermore, the hotter the surface, the 
further out the photospheric point will be, i.e.\ 
$\Delta r_2 > \Delta r_1$ in the above calculation. Because 
$P_{\rm ph,2}< P_{\rm ph,1}$, the temperature gradient 
$\nabla_2$ below an irradiated part  will be higher when setting 
$P_{\rm ph,2} = P_{\rm ph,1}$ rather than using the proper value of 
$P_{\rm ph,2}$. Therefore by assuming $P_{\rm ph,2}= P_{\rm ph,1}$ we 
overestimate the radiative flux on the irradiated part. We obtain the 
same result from using $\Delta r_2 = \Delta r_1$: because in reality 
$\Delta r_2 > \Delta r_1$, the temperature gradient (\ref{eq69}) on 
an irradiated part is lower than what our estimate with 
$\Delta r_2 = \Delta r_1$ yields. Therefore, our very simple one-zone 
model, i.e.\ Eq.~(\ref{eq70}) underestimates the blocking effect 
somewhat. Since, on the one hand, this deficit can easily be 
compensated for by slightly increasing the value of $b$, and since, 
on the other hand, the precise value of $b$ which is appropriate is 
not exactly determined within our model (we shall later take an 
average value determined from published opacity tables), we consider 
(\ref{eq70}) a fair approximation of the physical situation described. 
The really important aspect of our model is, however, that 
qualitatively it yields the correct behaviour of a stellar surface 
exposed to external irradiation, and that it is still simple enough 
to allow insight in the situation described. 

We can also compute $g=-dG/dx$. Differentiation of (\ref{eq70}) yields 
\bea
\label{eq71}
g&&=\left\{ \begin{array}{ccc}
\frac{\large nT^4_0 T^{n-1}_\irr}{\large nT^4_0 T^{n-1}_\irr + 
4 T^3_\irr (T^n_B-T^n_0)}\; ,& n=5-b\ne 0 \; \\
&&\\
\frac{\large T^4_0}{\large T^4_0 + 4T^4_\irr \ln (T_B/T_0)} \; ,& n=5-b = 0\;
\end{array}\right. ,
\eea
if $T_\irr < T_B$. For consistency with Eq.~(\ref{eq32}) we require 
$g=0$ if $T_\irr > T_B$. 

As can be seen from Eqs.~(\ref{eq70}) and (\ref{eq71}) the functions 
$G$ and $g$ depend only on two parameters characterizing the 
unirradiated star, namely on $T_0$ and $T_B$, and on the opacity law 
via $b$. While $T_0$ is a well-defined quantity, $T_B$ is not 
because in real stars the run of temperature $T$ with pressure $P$ is 
not as simple as the one assumed in our simple model (and sketched in 
Fig.~6). In particular, the transition from convective to radiative 
energy transport is smooth and does not occur at one particular point 
as we have assumed in our model. Since $T_B$ stands for the temperature 
at which this transition occurs, we determine $T_B$ by requiring that 
in a full   stellar model the ratio $F_\conv/F_\rad$ of convective 
flux $F_\conv$ to radiative flux $F_\rad$ reaches a prescribed value, 
say $F_\conv/F_\rad = k$. This is equivalent to the condition 
$\nabla_\rad = (k+1)\nabla$. Of course the choice of $k$ is somewhat 
arbitrary but a value $k\approx 1$ seems a natural choice. Our simple 
model is an acceptable description of the real situation only if for a 
given model $T_B(k)$ is sufficiently insensitive to $k$. In order to 
demonstrate that this is indeed the case we plot in Fig.~7 the run of 
$T_B$ as a function of the stellar mass of zero-age main-sequence stars 
with Pop.~I chemical composition (X = 0.70, Z = 0.02) for three different 
values of $k$, i.e. $k=1$ (full line), $k=0.5$ (dashed line) and $k=2$ 
(dotted line). Fig.~7 shows two important properties of low-mass stars: 
The first one is that a significant superadiabatic convection zone 
exists only in stars with a mass $M\ga 0.65\msol$. Below 
$M\approx 0.6\msol$ the stratification is essentially adiabatic up to 
the photosphere. This means that application of our simple one-zone 
model is restricted to stars in the mass range 
$0.65\msol\la M \la 1\msol$. The second property shown in Fig.~7 is 
that the run of $T_B(M)$ for different $k$ is qualitatively the same 
for all three values of $k$. This means that as long as a star has a 
significant superadiabatic convection zone the transition from 
convective to radiative energy transport occurs in a rather narrow 
temperature interval, thus justifying our simple approach. 

\begin{figure}[t] 
\plotone{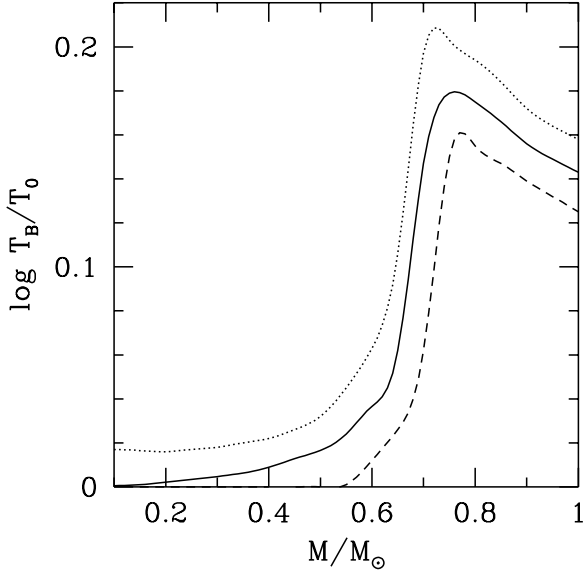}
\caption[]{Temperature $T_B$ of the point in the superadiabatic
convection zone of a main sequence star $F_\conv/F_\rad=k$, or
$\nabla_\rad=(k+1)\nabla$ for $k=0.5$ (dashed line), $k=1$ (full line),
and $k=2$ (dotted line) as a function of stellar mass. $T_B$ is measured
in units of $T_0$, the effective temperature of the unirradiated star. }
\end{figure}

\bigskip
\noindent{\it 5.2 Stars with a mass $M\la 0.6\msol$} 
\medskip

\stepcounter{subsection}
As Fig.~7 shows, $T_B \approx T_0$ in stars with a mass $M\la 
0.6\msol$. This means on the one hand that the total optical depth 
between the photosphere (at $T=T_0$) and the point where $T=T_B$ is 
small, in fact too small for our simple one-zone model, which assumes 
the diffusion approximation (Eq. \ref{eq63}), to be applicable. On 
the other hand, this means also that in these stars even in the 
photosphere a non-negligible fraction of the flux is transported by 
convection. Therefore, we must ask how irradiation changes the 
transported flux if the top of the (adiabatic) convection zone is at 
low optical depth. Because the convective flux is 
$F_\conv \sim (\nabla-\nabla_{ad})^{3/2}$ 
and the value of $\nabla$ is directly influenced by irradiation, this 
situation cannot be described by a simple model. Rather one ought to 
determine $\nabla$ by solving the full set of equations describing 
convective energy transport, i.e. in the simplest case the equations 
of mixing length theory. Considering the uncertainties inherent in 
current convection theories it is not obvious whether it is possible 
to make general statements about the functions $g$ or $G$. After all 
it is at least conceivable that already a small irradiating flux 
could result in a significant reduction of $(\nabla-\nabla_{ad})$ (which 
itself is a rather small number because convection is not far from 
adiabatic) and that therefore $\p F/\p F_\irr$ could attain a large 
negative value. However, as the following argument shows, there is a 
limit to how fast $F$ can drop in response to increasing $F_\irr$. 
If $F$ drops too strongly this results in an effective temperature 
$T_\irr < T_0$. This in turn means that the temperature gradient must 
be steeper than in the unirradiated star and, therefore, that $F>F_0$,
in contradiction to the starting assumption $\p F/\p F_\irr < 0$. In 
order to avoid this contradiction $T_\irr$ must not decrease with 
increasing $F_\irr$, i.e.\ $\p T_\irr/\p F_\irr \ge 0$, from which we 
immediately recover (\ref{eq33}), i.e.\ $g\le 1$. Because $g^\prime < 0$ 
(cf. 34), $g$ is maximal in the limit $F_\irr \to 0$. {\rev Therefore 
we need to determine $g(0)$ for the stars in question. For this we 
return to the one zone model (Sect. 5.1).} From Eq.~(\ref{eq71}) we 
find that $g(0)$ increases as $T_B/T_0$ decreases. In fact, in the 
limit $T_B = T_0$, this model yields $g(0)=1$. {\rev Thus} the closer 
the adiabatic convection zone reaches to the surface the more 
sensitive the star is to irradiation, i.e. the larger $g$. On the 
other hand, we know that $g\le 1$ in all cases. It is therefore 
plausible that for stars which are almost adiabatic up to the 
photosphere, i.e.\ $M\la 0.6\msol$, $g(0)\la 1$. As we shall see below 
this is confirmed by numerical computations. 

\begin{figure}[t] 
\plotone{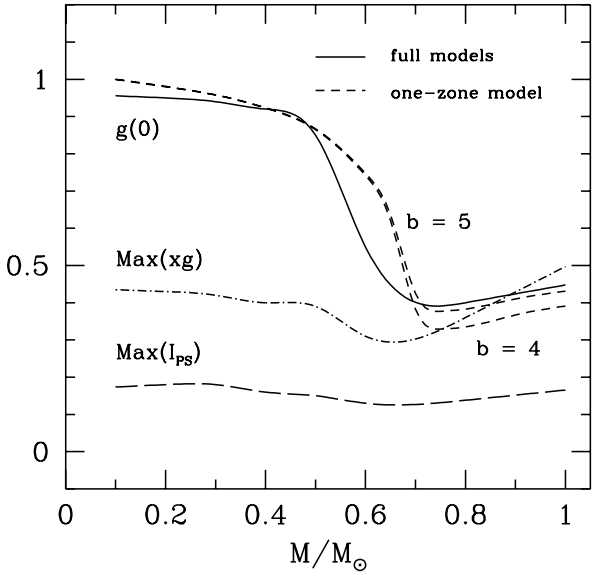}
\caption[]{The values of $g(x$=0) (see Eq. \ref{eq31}) for main
sequence stars as a function of stellar mass. Results from full
stellar models from Hameury and Ritter (1997) are shown as a full
line, results obtained from the one-zone model (Eq. \ref{eq71}) as
short-dashed  lines. For comparison we show also the run
of ${\rm Max}[xg(x)]$ (dash-dotted line) and ${\rm Max}(I_{PS})$ 
(long dashed line) derived from the results of Hameury and Ritter (1997).}
\end{figure}

\subsection{Results of numerical computations}  
Numerical computations of $g(x)$ and $G(x)$ {\rev have been} carried 
out by HR97 for low-mass main sequence stars of Pop.~I chemical 
composition using a full {\rev 1D} stellar structure code (Hameury 
1991), where the outer boundary condition was changed according to
Eqs.~(\ref{eq27}) and (\ref{eq28}) with $F_\irr=\mbox{const.}$ The
results relevant for this paper are shown in Fig.~8 and can be
summarized as follows:
\begin{itemize}
\item[a)] With $T_B(M)$ as shown in Fig.~7 and an appropriately 
chosen value for $b$, i.e.\ $b\approx 4-5$, typical for H$^-$ 
opacity, the predictions of our simple one-zone model are in good 
agreement with numerical results as long as $T_\irr < T_B$. In 
particular, the run of $g(0)$ shown in Fig.~8 as a full line for the 
numerical computations and as a short-dashed line for the one-zone 
model show that within its range of validity the latter reproduces 
the numerical results quite well. 
\item[b)] $g(0)\approx 1$ for stars with a mass $M\la 0.5\msol$, as we
have argued above (Sect.~5.2).
\item[c)] 
Max$(xg(x))$ $\la$ 0.5 over the whole mass range of interest, i.e. 
$0.1\msol \la M \la 1\msol$. This is shown as a dash-dotted line 
in Fig.~8.  
\item[d)] In the point source model (cf Sect.~4.2.2) the integral 
$I_{PS}$ in (\ref{eq59}) has a maximum which is smaller than 
$(1-f_s)$ Max$(xg(x))$. Max$(I_{PS})$ is shown in Fig.~8 as a long 
dashed line. As can be seen, Max$(I_{\rm PS}) \approx 0.20$.
\end{itemize}

\section{Secular evolution with irradiation} 
We resume the stability discussion of Sect.~4, but now 
with the knowledge of the function $g(x)$ which we have gained 
in Sect.~5. We shall restrict most of the following to the constant 
flux model (Sect.~4.2.1) and the point source model (Sect.~4.2.2). 
In addition, we shall assume that all the properties of a binary 
along a secular evolution without irradiation, in particular the 
functions $\Gamma$ and $\Lambda$ defined respectively in Eqs. 
(\ref{eq41}) and (\ref{eq46}), are known. Among the compact binaries 
we specifically discuss cataclysmic variables, and low-mass X-ray 
binaries. In particular, we wish to examine the following questions: 
a) which systems are stable (unstable) at 1) turn-on of mass transfer, 
or 2) during the secular evolution, against irradiation-induced 
runaway mass transfer, and b) what kind of evolution do systems 
undergo which are unstable? Because part of these questions have 
been dealt with extensively by King (1995), KFKR95, KFKR96, KFKR97
and MF98, we shall mainly be concerned with those aspects which have not 
already been treated in detail in the above papers.

\subsection{Cataclysmic variables}
In the following we set $M_c = M_{\rm WD}$ and $R_c = R_{\rm WD}$, 
where $M_{\rm WD}$ and $R_{\rm WD}$ are respectively the mass and 
the radius of the accreting white dwarf. Furthermore we shall 
restrict our discussion to CVs where the secondary is a low-mass 
main sequence star. Because this is the case for the vast majority 
of CVs this is not a very strong restriction. For determining the 
values of $\Lambda$ and $\Gamma$ we adopt the standard evolutionary 
scheme for CVs, i.e. the model of disrupted magnetic braking (e.g. 
King 1988 for a review) and use results of corresponding model 
calculations by KR92. 

\subsubsection{Stability of mass transfer at turn-on} 
Because $g(x)$ is maximal for $x=0$, systems are most susceptible 
to irradiation at turn-on of mass transfer. The relevant stability 
criterion for the constant flux model is (\ref{eq53}) and for the 
point source model Eq.~(\ref{eq60}). As we have already pointed out 
in Sect.~4 these conditions are in fact conditions on the 
efficiency factor $\eta$ because everything else is basically fixed. 
Given the masses of both binary components, i.e. $M_{\rm WD}$ and 
$M_s$, $R_{\rm WD}$ follows from the mass radius relation of white 
dwarfs, the secondary's radius $R_s$ from the mass radius relation 
of main sequence stars, or from the evolutionary history. The orbital 
separation $a$ follows from Roche geometry via $R_s$ and the mass 
ratio, $\zeta_S$ and ${\cal F}$ from the secondary's internal 
structure, and finally $\zeta_R$ again from the mass ratio. Thus 
$\Gamma$ (Eq. \ref{eq46}) is uniquely determined by $M_{\rm WD}$ 
and $M_s$. Given the function $g(x)$, and in particular $g(0)$, 
condition (\ref{eq53}) is one for $\eta s$, and condition  
(\ref{eq60}) one for $\eta$ only. Taking as an example $g$ from 
our one-zone model with $b=4$, $s=0.5$ and $T_B/T_0$ for $k=1$ 
from Fig.~7, we can plot a line for a given $\eta$ in the $ 
M_s$-$M_{\rm WD}$ plane along which mass transfer is marginally 
stable. This is shown in Fig.~9 for various values of $\eta$. 
The figure is to be read as follows: a parameter combination 
($M_{\rm WD}$, $M_s$) corresponds to a point in Fig.~9. If that 
point lies above the line corresponding to {\rev a given} value of 
$\eta$, mass transfer is unstable at turn-on. What we can infer 
from Fig.~9 is that for typical WD masses $M_{\rm WD} \ga 0.5\msol$, 
condition (\ref{eq53}) is violated for surprisingly small values of 
$\eta$, i.e.\ $\eta \la 0.1$. This is the case, in particular, if 
$M_s \ga 0.6\msol$. If we take instead of (\ref{eq53}) the condition 
of the point source model (\ref{eq60}), $s$ is no longer a parameter. 
Taking again the same $g$ as above (one-zone model with $b=4$) the 
result is qualitatively the same. The main difference is that the 
value of $\eta$ necessary for marginal stability needs to be larger 
by about a factor of 2. 

\begin{figure}[t] 
\plotone{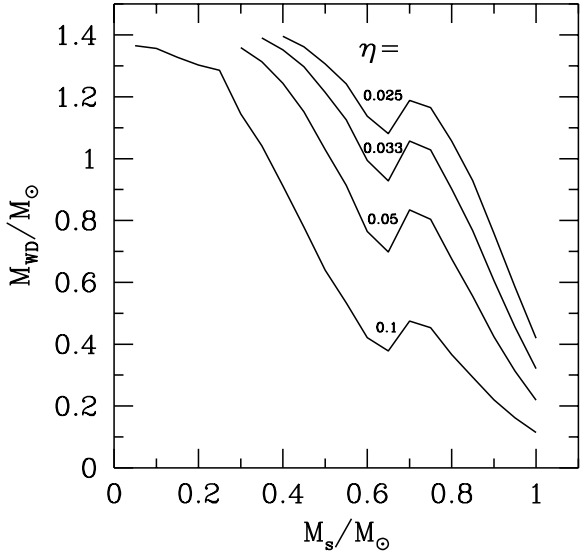}
\caption[]{Contour lines in the $M_{\rm WD}-M_s$-plane along which 
$\zeta_\irr(F_\irr=0)=1$ for different values of $\eta$. Systems above
(below) a particular line are unstable (stable) against irradiation-induced
runaway mass transfer at the turn-on of mass transfer.}
\end{figure}

The trends seen in the curves of Fig.~9 are easily explained: they 
reflect the run of $\Gamma (M_{\rm WD},M_s)$ (cf Eq.~46). The smaller 
$\Gamma$ the less stable mass transfer. Because $\Gamma$ is 
proportional to $R_{\rm WD}/M_{\rm WD}$ and ${\cal F}^{-1}$, and 
$R_{\rm WD}/M_{\rm WD}$ is a steeply decreasing function of 
$M_{\rm WD}$ whereas ${\cal F}^{-1} \sim M_{\rm ce}/M_s$ scales 
roughly as the relative mass of the convective envelope (cf. Fig.~1, 
dashed line) and thus decreases strongly with $M_s$, mass transfer is 
more likely to be unstable (stable) the higher (lower) $M_{\rm WD}$ 
and $M_s$. The other factors entering $\Gamma$, i.e. 
$(\zeta_S-\zeta_R)$, $M_s/R_s$ and $(a/R_s)^2$ are of comparatively 
minor importance. The fact that the curves in Fig.~9 are not 
monotonic results from the steep increase of $T_B/T_0$ with stellar 
mass near $M=0.65\msol$ (cf. Fig.~7). 

Because $g^\prime(x)<0$, systems in which mass transfer is stable 
at turn on will also be stable against irradiation-induced runaway 
mass transfer for any finite value of $F_\irr$, i.e. $\dm$. The 
opposite, however, is not true: not all systems which are unstable 
at turn-on will be so for the secular mean mass transfer rate. 

\begin{figure}[t] 
\plotone{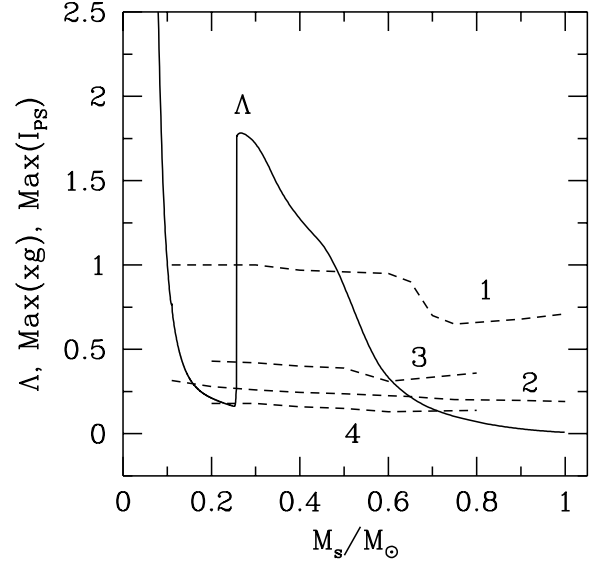}
\caption[]{The functions $\Lambda$ (Eq. \ref{eq41}) (full line),
${\rm Max}[xg(x)]$ and ${\rm Max}(I_{\rm PS})$ according to the
one-zone model (dashed lines 1 and 2 respectively), and ${\rm
Max}[xg(x)]$ and ${\rm Max}(I_{\rm PS})$ derived from the results of
Hameury and Ritter (1997) (dashed lines 3 and 4 respectively) 
as a function of the secondary's mass along a standard evolution of a CV
with $M_{\rm WD}=1\msol$ and $M_{s,i}=1\msol$ calculated by KR92. }
\end{figure}

\subsubsection{Stability of secular evolution} 
The appropriate stability criterion is (\ref{eq41}) in its most 
general form, (\ref{eq52}) for the constant flux model and 
(\ref{eq59}) for the point source model. Again, the left-hand side 
$\Lambda$ of these criteria is fully determined by the adopted model 
of secular evolution, whereas the value of the corresponding 
right-hand sides still depends on $\eta$. As we have stressed in 
Sect.~4, we can give a sufficient criterion for stability of mass 
transfer because the function $xg(x)$ has a maximum. The corresponding 
criterion is (\ref{eq44}) in its most general form, 
\be
\Lambda > 2s \; \mbox{Max} (xg(x))\label{eq72}
\ee
for the constant flux model, and
\be
\Lambda > 
\mbox{Max} (I_{\rm PS})\label{eq73}
\ee
for the point source model. Conditions (\ref{eq72}) and (\ref{eq73}) 
can be read off Fig.~10 where we plot $\Lambda$ as a function of 
secondary mass $M_s$ (full line) along a standard secular evolution 
of a CV with $M_{\rm WD} = 1\msol$ and an initial secondary mass 
$M_{s,i} = 1\msol$. The evolutionary data are taken from KR92. Into 
the same figure we plot Max$(xg(x))$ $(=2s$ Max$(xg(x))$ for $s=1/2$) 
and Max$(I_{\rm PS})$ using results by HR97 for $g(x)$ (dashed lines). 
As can be seen the values of Max$(xg(x))$ and Max$(I_{\rm PS})$ are 
almost independent of secondary mass, i.e. of the orbital period. For 
CVs above the period gap, i.e. with $M_s \ga 0.25\msol$, $\Lambda$ 
increases very steeply with decreasing secondary mass. This behaviour 
is mainly due to two factors, first to 
${\cal F}^{-1}\sim M_{\rm ce}/M_s$ which increases strongly with 
decreasing mass, and second to $\tau_{\rm KH}$ which increases 
strongly towards lower masses mainly because of the mass 
luminosity relation $L\sim M^3$ of low-mass main sequence stars. So, 
what $\Gamma$ essentially represents is the thermal inertia of the 
convective envelope. As can be seen from Fig.~10, the intersection 
of $\Lambda$ with Max$(xg(x))$ or Max$(I_{\rm PS})$ is near $M_s = 
0.7\msol$. Because $\Lambda$ increases so steeply, small changes in 
Max$(xg(x))$ or Max$(I_{\rm PS})$ do not yield a significantly 
different result.

When a CV approaches the period gap, i.e. the secondary becomes 
fully convective at $M_s \approx 0.25\msol$, the system detaches and 
$\Gamma \to 0$ because $\tau_{M_s}\to \infty$. When mass transfer 
resumes $\Lambda$ is smaller by typically a factor 10 than 
immediately above the gap, reflecting the reduced angular momentum 
loss below the gap and the fact that at turn-on $R_s = R_0$ 
(cf. \ref{eq41}). $\Lambda$ starts increasing again with further 
decreasing mass because the Kelvin-Helmholtz time $\tau_{\rm KH}$ 
becomes very long as the secondary becomes degenerate.

Because $\Lambda$ does not depend strongly on the mass of the white  
dwarf we can generalize the result found from Fig.~10: In the 
framework of standard CV evolution, i.e. the model of disrupted 
magnetic braking, CVs are stable against irradiation-induced runaway 
mass transfer when the mass of the secondary star is $M_s \la 
0.7\msol$. If, on the other hand, $M_s\ga 0.7\msol$, a system can be 
unstable but need not be so, subject to the value of $\eta$. 

We note, however, that by invoking substantial {\rev consequential
angular momentum loss (CAML), i.e. angular momentum loss which 
depends explicitly on the mass transfer rate (see King and Kolb 1995 
for a discussion of CV evolution with CAML)}, the mass range over 
which systems can be unstable is much larger. This has already been 
pointed out by KFKR96 and confirmed in computations by MF98.

As can be seen in Fig.~10 there is also a slight chance that CVs 
immediately after turn-on below the period gap are unstable. 
According to the constant flux model (Eq.~72), some CVs could be 
unstable, according to the point source model, which is more 
realistic but still very optimistic, the stability criterion 
(\ref{eq73}) is violated only marginally if at all. Therefore, our 
conclusion is that CVs below the period gap are very probably 
stable, at least in the framework of standard CV evolution.
 
Examining the factors which determine the value of $\Lambda$ 
(Eq.~41) we see that within a given evolutionary model all 
factors are determined. $\zeta_S$, $\tau_{\rm KH}$ and ${\cal F}$ 
depend only on the mass of the secondary and its evolutionary 
history, $\tau_{M_s}$ on the adopted rate of angular momentum loss, 
$\zeta_S-\zeta_R$ and the evolutionary history (via 
$(\p\ln R_s/\p t)_{\rm th}$ 
in (\ref{eq3})), $\zeta_R$ on the mass ratio and $R_s/R_0$ on the 
evolutionary history (e.g. Stehle et al. 1996). We see also that 
making $\Lambda$ smaller (in order to get the instability for 
lower secondary masses) is possible only by either increasing 
$\tau_{M_s}$, i.e. lowering the mean mass transfer rate 
$\langle \dot M_s\rangle$, or by increasing $\zeta_R$. The former 
is practically impossible without upsetting the standard evolutionary 
paradigm for CVs, i.e. the period gap model which requires that above 
the period gap ($M_s\ga 0.25\msol$, $P\ga 3^h$) 
$\tau_{\rm KH}/\tau_{M_s} \ga 5$ (e.g. Ritter 1984, King 1988; KR92; 
Stehle et al. 1996). Increasing $\zeta_R$, on the other hand, is 
possible only if a system experiences significant CAML.

\subsection{Low-mass X-ray binaries} 
At first glance one might suspect that in LMXBs irradiation of the 
secondary represents a much larger threat for the stability of mass 
transfer than it does in CVs. {\rev However, for the following 
reasons this is very probably not the case}: first we note 
the observational fact that very few of the LMXBs show X-ray eclipses. 
This has been interpreted as a consequence of the large vertical scale 
height of the X-ray irradiated accretion disk. This in turn allows the 
secondary to stay permanently in the disk's shadow. If this is the 
case, none or at most a small part of the secondary's surface is 
directly exposed to the accretion light source. {\rev Second, indirect 
illumination of significant parts of the donor (the high latitude 
regions or part of the back side) is ruled out because this would 
require a very extended scattering corona indeed, with a typical size 
of the scattering sphere (at optical depth $\tau \sim 1$) of order or 
larger than the donor star. As a consequence one would expect X-ray 
eclipses to occur much more frequently 
than they are actually observed. Third, heat transport by currents from 
the hot, illuminated to the cool parts in the X-ray shadow are probably 
also negligible. This is because in cool stars with a deep convective 
envelope the superadiabatic convection zone isolates the interior from 
the surface. It is itself unable to transport significant amounts of 
heat because of its small heat capacity and the fact that the thermal 
time scale is much shorter than the time scale of circulation.  

Heat transport by currents caused by hydrostatic disequilibrium is, 
however, of importance in stars with a radiative envelope. Effects of 
this are e.g. seen in the X-ray binaries HZ Her and V1033 Sco (see e.g. 
Shahbaz et al. (2000), and references therein). 

Because neither indirect llumination nor heat advection can contribute 
significantly to the blocking of the stellar flux,} the integral $I$ 
on the right-hand side of (\ref{eq41}) is much smaller than for a 
comparable CV, despite the fact that in a LMXB there is potentially 
much more energy available for irradiating the secondary. The fact 
that $xg(x)$ has a maximum {\rev at $x \approx 1$} is the reason why 
even large irradiating fluxes do not help. {\rev Rather,} optimal 
irradiation is achieved if as large a fraction as possible of the 
secondary's surface is irradiated with a flux such that $xg(x)$ is 
near its maximum, i.e. if $x\approx 1$. This is clearly not the case 
in LMXBs. Not only is the surface fraction {\rev which is directly 
irradiated} small, worse, where the surface of the secondary is 
directly exposed to the X-ray source, the associated flux is large, 
i.e. $x\gg 1$ unless $\eta$ is assumed to be very small. The latter 
is very unlikely considering the fact that most of the accretion 
luminosity emerges in form of rather hard X-rays. Thus, unless the 
secular evolution of LMXBs is totally unlike that of CVs as far as 
the nature of the secondary  and the typical mass transfer rates are 
concerned, the value of $\Lambda$ is virtually the same as for CVs 
but, as explained above, the right-hand side of (\ref{eq41}) is much 
smaller than in CVs. Therefore we conclude that very probably LMXBs 
are stable against this type of irradiation-induced runaway mass 
transfer. As has been noted by KFKR97, systems in which accretion is 
intermittent rather than continuous, i.e. transient LMXBs, are even 
more stable.

\subsection{Evolution of unstable systems} 
Let us now discuss briefly the evolution of systems (i.e. CVs) in 
which the stationary mass transfer given by (\ref{eq3}) is unstable, 
i.e. systems for which the stability criterion (\ref{eq41}) or a 
special form thereof (Eqs. \ref{eq52} or \ref{eq59}) is violated. 
From the fact that $g^\prime (x)<0$ we know that mass transfer is 
then already unstable at turn-on. Therefore, when mass transfer 
turns on, the mass transfer rate increases, and because (\ref{eq41}) 
is violated, it does not settle at the secular mean 
$\langle-\dm\rangle$ given by (\ref{eq3}). However, because the 
thermal relaxation caused by irradiation saturates both in amplitude 
and with time (see our discussion in Sect.~3, in particular 
Eqs.~(\ref{eq20}) and (\ref{eq23}), and Fig.~4) the mass transfer 
rate does not run away without bound. Rather, there is an upper limit: 
with (\ref{eq20}) we obtain for the maximum mass transfer rate 
\bea
\label{eq74}
\mbox{Max}(-\dm)&\approx& \langle-\dm\rangle + \frac{M_s}{(\zeta_S-\zeta_R)} \; 
\mbox{Max} \left\{\frac{s_\eff}{\tau_{\rm KH}}\, {\cal F}\right\} \nonumber\\
&\approx& \langle -\dm\rangle + \frac{s}{\zeta_S-\zeta_R} 
\frac{M_s}{\tau_{\rm KH}} \; {\cal F}\quad.
\eea
Because mass transfer can be unstable against irradiation only if 
$M_s \ga 0.7\msol$, i.e.\ when the secondary has a relatively thin 
convective envelope and thus ${\cal F}$ is large (cf. Fig.~1), the 
maximum contribution of irradiation to mass transfer can be several 
times ($s{\cal F}$ times) the thermal time scale mass transfer rate 
$M_s/\tau_{\rm KH}$. Thus, for such systems 
Max$(-\dm) \gg\langle-\dm\rangle$. 

After having reached the peak mass transfer rate, mass transfer 
cannot continue at that rate. Rather it must decrease with time 
for two reasons: first, thermal relaxation saturates on the time 
scale given by (\ref{eq23}), i.e. $(\p\ln R/\p t)_\irr$ decreases 
on that time scale. Second, because mass transfer occurs at a rate 
above the secular mean, the binary system is driven apart, i.e. 
$(d\ln R_R/dt)$ $>$ $(d\ln R_s/dt)$. Both effects result eventually 
in the termination of mass transfer. The system becomes slightly 
detached, the secondary, in the absence of irradiation, shrinks. 
However, because of the absence of mass transfer the contraction of 
the system due to angular momentum loss is fast enough to catch up, 
so that mass transfer resumes and the cycle repeats again. In other 
words: if a system is unstable at the secular mean mass transfer 
rate it must undergo a limit cycle in which phases of enhanced, 
irradiation-driven mass transfer alternate with phases of very low 
or no mass transfer. The conditions for the occurrence of mass 
transfer cycles in semi-detached binaries have been investigated in 
more detail and using more general principles (in the framework of 
non-linear dynamics) by King (1995), KFKR95, KFKR96 and KFKR97. 
Their main result is that mass transfer cycles driven by radius 
variation of the secondary can only occur if 
$(\p\ln R_s/\p t)_{\rm th} +(\p\ln R_s/\p t)_\nuc$ in (\ref{eq1}) 
depends explicitly on the {\it instantaneous} mass transfer rate. 
The only plausible mechanism providing such a dependence is 
irradiation of the donor star by radiation generated through 
accretion, i.e. the situation we are studying in this paper. The 
necessary criterion for the occurrence of mass transfer cycles 
found by these authors is identical to what we have found here, 
namely the violation of (\ref{eq41}). In the framework of a linear 
stability analysis of mass transfer, with which we, King (1995), 
KFKR95, KFKR96 and KFKR97 have been concerned so far, we can not 
{\rev calculate} the long-term evolution over time scales 
$\tau_{M_s}$ of systems under the irradiation instability. For this 
the full set of equations describing mass transfer and stellar 
structure under irradiation have to be solved. Results of such 
calculations and computational details will be presented in the 
following section. Because similar computations have already been 
done by HR97 and MF98, we shall concentrate here on aspects which 
have not been dealt with in detail by HR97 and MF98, but are, in our 
opinion, important for better understanding of the evolution under 
the irradiation instability.

\section{Secular evolution with irradiation: numerical computations} 

\subsection{Computational techniques} 
To compute the secular evolution of a compact binary with a 
low-mass donor star we have used the bipolytrope programme described 
in detail in KR92. With respect to the procedure described in KR92 we 
have, however, implemented two modifications in order to allow for a 
proper treatment of irradiation and its consequences for the secular 
evolution.

First, we compute the mass transfer rate explicitly rather than by 
using Eq.(\ref{eq3}) as in KR92, i.e.\ we adopt the following 
prescription (e.g.\ Ritter 1988): 
\be
-\dm = \dot{M}_0 \exp \eck{-\frac{R_R - R_s}{H_P}} \label{eq75}
\ee 
Expressions for {\rev the photospheric pressure scale height} $H_P$ 
and the factor $\dot M_0$ in terms of stellar parameters of the donor 
star and binary parameters are also given in Ritter (1988). In the 
context of this paper the donor star is always a low-mass main 
sequence star. For such stars both, $H_P$ and $\dot M_0$ are only 
weak functions of the stellar mass and have typical values 
$H_P/R_s \approx 10^{-4}$ and $\dot M_0 \approx 10^{-8}\msol$yr$^{-1}$. 
The reason for using (\ref{eq75}) instead of (\ref{eq3}) is that the 
latter is valid only for stationary mass transfer. {\rev However, 
this} is not a good approximation when a system evolves through mass 
transfer cycles, because such cycles proceed unavoidably through 
phases of non-stationary mass transfer. In order to simplify the 
computation of $-\dm$ we used fixed values for 
$H_P$ and $\dot M_0$, i.e.\ $H_P = 10^{-4}R_s$ and 
$\dot M_0 = 10^{-8} \msol$yr$^{-1}$. Furthermore we used (\ref{eq75}) 
also when $-\dm > \dot M_0$, i.e. when the donor star overfills its 
critical Roche volume. (\ref{eq75}) is still a reasonable 
approximation if $R_s - R_R <$ few $H_P$, i.e. 
$-\dm \la 10^{-7}\msol$yr$^{-1}$ (see e.g.\ Kolb and Ritter 1990) 
which is adequate for our purposes. 

When computing the mass transfer rate from (\ref{eq75}) we use 
for $H_P$ and $\dot M_0$, both of which depend on the effective 
temperature, the value of $T_\eff$ on the unirradiated part of 
the star, i.e. $T_\eff = T_0$. There are two main arguments for 
this choice: first, in disk-accreting systems the inner Lagrangian 
point $L_1$ is in the shadow of the disk and the cooling time of 
gas in the superadiabatic layer flowing from the irradiated parts 
towards $L_1$ is short compared to the flow time in the shadow 
region. Second, if we use $T_\eff = T_\irr$ instead of $T_\eff = 
T_0$, $H_P$ and $-\dot M_s$ would react practically instantaneously 
and with a large amplitude to irradiation giving, in turn, rise to a 
runaway of the mass transfer rate on an extremely short time scale 
(order of hours). This is obviously not what happens in the systems 
we do observe.

Second we use either Eq.~(\ref{eq54}) in the constant flux 
model or Eq.~(\ref{eq61}) in the point source model of irradiation 
in place of the usual Stefan-Boltzmann law as one of the outer 
boundary conditions for the donor star model. For the calculations 
presented below we have used the one-zone model described in 
Sect.~5 rather than results obtained from full stellar models by 
HR97 described earlier. Specifically, $T_\irr$ in (\ref{eq54}) or 
(\ref{eq61}) was computed by solving (\ref{eq70}) in which $F_\irr$ 
is determined for an adopted value of $\eta$ from (\ref{eq24}) with 
$h(\vart,\varp)$ taken respectively from (\ref{eq50}) and 
(\ref{eq51}) in the case of the constant flux model, and from 
(\ref{eq56}) in the case of the point source model. $T_B$ in 
(\ref{eq70}) is taken from the numerical results shown in Fig.~7. 
Specifically, we have used $T_B (M_s)$ for $k=1$. Furthermore, in 
order to determine $T_\irr$ we need to specify $b$, i.e. the 
temperature exponent of the photospheric opacity law (\ref{eq10}). 
For most of our experiments we have used $b=4$ which is adequate 
for H$^-$ opacity. 

The diagnostic quantities $\zeta_\irr$ and $s_\eff$ are calculated 
from (\ref{eq45}) and respectively (\ref{eq55}) or (\ref{eq62}) with 
the appropriate choice of $h(\vart,\varp)$, i.e.\ Eqs.~(\ref{eq50}) 
or (\ref{eq56}), and $g(\vart,\varp)$ from (\ref{eq71}).

\subsection{Computational limitations} 
By using the bipolytrope approximation for describing the internal 
structure of the donor star, we can only deal with chemically 
homogeneous stars which have a radiative core and a convective 
envelope or are fully convective, i.e. with low-mass $(M_s\la 1\msol)$ 
zero age main sequence stars. Thus, we are unable to address 
chemically evolved stars, in particular subgiants. Giants, on the 
other hand, have been dealt with approximately by KFKR97. We note 
also that chemically evolved donors among CVs might be more common 
than hitherto thought (Kolb and Baraffe 2000; Ritter 2000; Baraffe 
and Kolb 2000) and therefore deserve further study. For our 
computations, we have adopted Pop.~I chemical composition. In the 
context of the bipolytrope approximation, the chemical composition 
is of relevance mainly for determining the appropriate gauge 
functions, i.e. polytropic index $n_1$ of the radiative core and 
entropy jump $h$ in the surface layers as a function of stellar mass 
(for details see KR92). Even when adopting the appropriate gauge 
functions, we know that the value of $\zeta_S$ computed in the 
bipolytrope approximation is smaller than what one would obtain for 
a full stellar model if $M\ga 0.6\msol$. Because of this, in the 
bipolytrope approximation binary systems are more likely to be 
unstable against irradiation than they are in reality if 
$M_s\ga 0.6\msol$. 
 
We have used the one-zone model described in Sect.~5. As we have 
discussed there, application of this model is practically restricted 
to stars with a mass $M_s \ga 0.6-0.7\msol$. We have, however, 
carried out a few calculations of systems with a smaller secondary 
mass. In those cases we have assumed $g(x) = 1$ if $0\le x\le 1$ and 
$g=0$ otherwise. This corresponds to what (\ref{eq71}) yields in the 
limit $T_B \to T_0$ and approximates results of numerical calculations, 
at least for small fluxes, i.e. $x<1$, reasonably well.
 
Our calculations contain a number of free or at best not 
well-constrained parameters. To mention just the most important ones: 
in the constant flux model we already make a very simplifying 
assumption about the function $h(\vart,\varp)$, i.e.\ the 
irradiation geometry. This assumption results in {\rev the} free 
parameter $s$. In addition, we have the efficiency factor $\eta$, 
a parameter about which we know little beyond the fact that probably 
$0<\eta\la 1$. Using our one-zone model (Sect.~5) introduces 
furthermore the parameters $T_B$ and $b$, both of which can, however, 
be fixed reasonably well by comparison with full stellar models. In 
the point source model we do not need to specify $s$. But all the 
other parameters, i.e. $\eta$, $T_B$ and $b$ remain in the problem. 
{\rev By using} numerical results for the functions $G(x)$ and $g(x)$ 
would {\rev get} rid {\rev of} the parameters $T_B$ and $b$. But we 
would still be left with specifying $h(\vart,\varp)$ and $\eta$. 
On top of all that we have also a number of input parameters and 
functions which are already needed for computing a secular evolution 
without irradiation. Apart from the binary's initial parameters the 
most important of those are the angular momentum loss rate and 
parameters arising from assumptions about mass and consequential 
angular momentum loss from the binary system. 

{\rev The purpose of the following computations, going beyond the
linear stability analysis, is to illustrate the temporal evolution
under the irradiation instability and a number of its specific
properties which we have uncovered in the foregoing discussion.} 

As we have demonstrated in the previous section the irradiation 
instability is more likely of relevance for CVs than for LMXBs. 
Consequently, in the examples below we have assumed that the 
accretor is a white dwarf with a mass $M_{\rm WD} = 1\msol$ and a 
radius according to the mass radius relation (e.g.\ Nauenberg 1972) 
of $R_{\rm WD} = 5~10^8$cm. We assume that during the secular 
evolution the mass of the white dwarf remains constant, i.e.\
$\langle\dot M_{\rm WD}\rangle=0$, and that on average the transferred 
matter leaves the system with the specific orbital angular momentum 
of the white dwarf. Other parameters characterizing the three examples 
which we shall discuss subsequently in detail are listed in Table~2.

\begin{table}
\caption[]{Parameters and model assumptions used for calculating 
the evolution shown in Figs. 11-13. Parameters common to all three 
examples are: $M_{\rm WD}=1\msol$, $R_{\rm WD}=5\, 10^8$cm, 
$H_p/R=10^{-4}$, $\dot{M}_0 = 10^{-8}\msol {\rm yr}^{-1}$; 
irradiation model: one-zone model with $b=4$; irradiation geometry: 
constant flux over surface fraction $s=0.5$.}
\begin{center}
\begin{tabular}{|l|l|l|l|}\hline
parameter & \multicolumn{3}{c|}{value} \\
&         Fig. 11 & Fig. 12 & Fig. 13 \\ \hline
$M_{s,i}(\msol)$ & 0.80 & 0.70 & 0.25\\
$\eta$               & 0.035 & 0.035 & 0.20\\
$\tau_J(yr)$        & $1.27\, 10^8$ & $1.27\, 10^8$ & $2.95\, 10^9$\\ \hline
\end{tabular}
\end{center}
\end{table}

\begin{figure}[t] 
\plotone{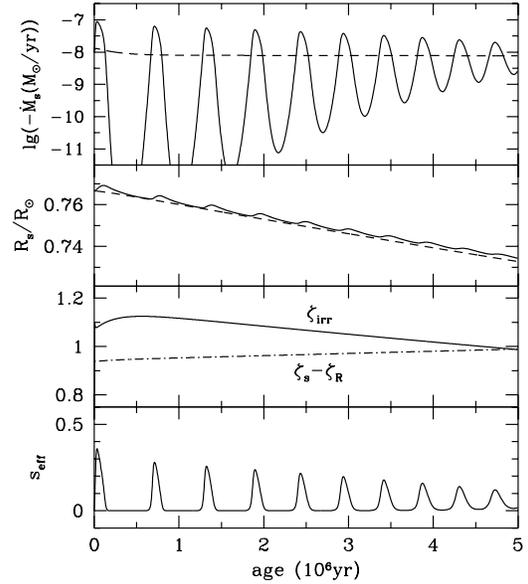}
\caption[]{Evolution of a CV with an initial secondary mass
$M_{s,i}=0.8\msol$ through mass transfer cycles computed by adopting 
the constant flux model with $s=0.5$ and the one-zone model with $b=4$.
Further parameters and model assumptions are listed in Table 2. Shown 
as a function of time are: in the top frame the mass transfer rate 
with and without taking into account the effect of irradiation (full 
and dashed line respectively); in the second frame the secondary's 
radius with and without taking into account the effect of irradiation 
(full and dashed line respectively); in the third frame 
$\zeta_\irr$ according to Eq. (\ref{eq40})(full line) and 
$\zeta_S-\zeta_R$ (dashed line); in the bottom frame $s_\eff$ 
according to Eq. (\ref{eq55}).}
\end{figure}

\subsection{Numerical results}

For the first example of a secular evolution with irradiation, 
results of which are shown in Fig.~11, we have adopted the constant 
flux model with $s=0.5$, an initial secondary mass $M_{s,i} = 0.8\msol$ 
and loss of orbital angular momentum on a constant time scale 
$\tau_J = -J/\dot J = 1.27~10^8$yr derived from the Verbunt and 
Zwaan (1981) prescription for magnetic braking with $f_{\rm vz}=1$. 
The value $\eta = 0.035$ was chosen such that on the one hand 
initially $\zeta_\irr > \zeta_S-\zeta_R$ and on the other 
$T_\irr < T_B$ at all times. The top panel of Fig.~11 shows the run 
of the  mass transfer rate with irradiation (full line) and without 
(dashed line), the second the evolution of the secondary's radius 
with irradiation (full line) and without (dashed line). In the third 
panel we show the run of $\zeta_\irr$ (full line) and of 
$\zeta_S-\zeta_R$ (dash-dotted line) with time. Finally the bottom 
panel shows the run of $s_\eff$ with time. As was to be expected for 
a system in which (initially) $\zeta_\irr > \zeta_S -\zeta_R$ mass 
transfer evolves through cycles with at least initially large 
amplitude. The amplitude of the radius variations, on the other hand, 
are small, a consequence of the small value of $H_P/R = 10^{-4}$. 
According to (\ref{eq75}) we have 
$\Delta \log R_s = H_P/R\; \Delta \log \dm$. We see also that during 
the mass transfer peaks a significant fraction of the stellar 
luminosity is blocked. Because $\eta$ has been chosen such that 
always $T_\irr < T_B$ we always have $s_\eff < s$. What is 
immediately apparent from this calculation is that though the 
system evolves through mass transfer cycles, these are damped on a 
rather short time scale, i.e. a time scale much shorter than $\tau_J$. 
{\rev The reason for this is that $\zeta_{\rm irr}$ decreases rapidly 
with time (mass of the donor). Eventually 
$\zeta_\irr < \zeta_S - \zeta_R$ and mass transfer becomes stable. 
This is mainly a consequence of the 
increasing thermal inertia of the convective envelope, i.e. of a 
decrease of ${\cal F}/\tau_{\rm KH}$ with decreasing donor mass. 
The damping of the oscillations} could only be overcome if at the 
same time $g$ increases sufficiently strongly with decreasing mass. 
We shall show below that in a restricted mass range this is {\rev 
indeed} possible. 

Given the results of the above example we now ask to what extent 
they are representative, i.e. whether the qualitative behaviour 
depends strongly on the adopted model parameters or not. With this 
end in view we have carried out numerous experiments, the results 
of which we shall now discuss.

Working with the one-zone model (Sect.~5) we ask first how the 
above results change with the parameter $b$, i.e. the adopted 
photospheric opacity. Inspection of Eq.~(\ref{eq71}) shows that 
for given values of $T_B$ and $T_\irr$ $g(0)$ increases with $b$. 
This means that at least for small fluxes the donor star is 
more sensitive to irradiation for larger $b$. The reason for this 
is easy to understand: the larger $b$ the more pronounced the 
increase of the optical depth in the superadiabatic layer in 
response to irradiation, i.e. of the average temperature, and 
thus the more effective the blocking of the energy outflow from 
the adiabatic interior. Thus, increasing (decreasing) $b$ above 
(below) the value $b=4$ we have used in the example shown in 
Fig.~11 results in more (less) pronounced mass transfer cycles. 
The time scale on which the mass transfer oscillations are damped 
remains, however, essentially uneffected by changes of $b$, 
reflecting the fact that ${\cal F}/\tau_{\rm KH}$ does not depend 
on $b$. 

Next we compare the results obtained with the constant flux model 
(Fig.~11) with those obtained with the point source model. The 
results of a run with the latter model and parameters identical to 
those used for producing Fig.~11 (except of $s$ which is not a 
free parameter in this model) are qualitatively very similar to 
those found in Fig.~11. The amount of stellar flux blocked during 
a mass transfer peak is, however, systematically smaller (by about 
a factor of two) than in the constant flux model. The main reason 
for this is that the irradiated area on the donor 
$s_{\rm PS} =0.5 (1-f_s) = 0.32$ is smaller by about a factor of 
1.6 than what we have assumed in the constant flux example, i.e. 
$s=0.5$. Therefore, in order to achieve the same effect as in the 
constant flux model with $s=0.5$, $\eta$ in the point source model 
needs to be increased by about a factor of two. Otherwise the 
results obtained from the two models are very similar. Because of 
this and because the constant flux model is {\rev computationally 
much less demanding} we have performed most of our simulations 
with that model. 

Next we examine briefly the dependence on the initial mass of the 
donor star. From our extensive discussions in Sects.~4 and 6 we 
know already that below a critical mass which is between about 0.6 
and $0.7\msol$, depending on the adopted models, systems following 
a standard CV evolution are stable against irradiation. 

Adopting the constant flux model and the one-zone model, a system 
with an initial secondary mass of $0.7\msol$ is still unstable 
(whereas {\rev with} the point source model {\rev and} a more 
realistic $g(x)$ such a system would be stable). Results of an 
evolution with $s=0.5$ and $\eta=0.04$ and the remaining parameters 
as in Fig.~11 (cf. Table~2) are shown in Fig.~12. As can be seen 
this evolution differs in several respects from the one shown in 
Fig.~11. {\rev Initially} the amplitude of the mass transfer cycles 
increases with time. After only a few cycles the peak mass transfer 
rate is so high that the irradiation effect saturates, i.e.\ 
$T_\irr > T_B$ and $s_\eff = s = 1/2$. Although the one-zone model 
used here does not apply when $T_\irr$ approaches $T_B$, the main 
effect of saturation can be modelled anyway by setting $g=0$ if 
$T_\irr \ge T_B$. The qualitative behaviour obtained in this way 
remains the same as if a more realistic and smooth function $g(x)$ 
was used. After an initial phase of increasing amplitudes they later 
start decreasing and die out very rapidly after about $4~10^7$yr. 
{\rev This behaviour can be understood as follows: We have already
pointed out above that} increasing amplitudes of the cycles can only 
be expected if with decreasing mass $g(x)$ increases fast enough. 
This is exactly what happens in the evolution shown in Fig.~12. The 
fast increase of $g(x)$ is the result of the fast decrease of 
$T_B/T_1$ when going from $M_s \approx 0.7\msol$ to 
$M_s \approx 0.6\msol$ (see Fig.~7). Below $M_s \approx 0.6\msol$, 
$g(x)$ does no longer change much with mass. As a result, the mass 
transfer cycles are then damped because of the secular increase of 
$\tau_{\rm KH}/{\cal F}$. When the system eventually becomes stable, 
irradiation is still important in blocking the energy outflow. As 
can be seen from the bottom panel of Fig.~12, $s_\eff$ is of the 
order 0.37 after the system has stabilized. 

Going to even lower initial secondary masses, mass transfer cycles 
cannot occur unless either $\zeta_S - \zeta_R$ or 
\break $\langle\-\dm\rangle$ is much smaller than in a standard CV 
evolution (cf.\ our discussion in Sect.~6.1). For illustrating 
this we show in Fig.~13 the results of a calculation for which we 
have assumed $M_s =0.25\msol$, $\eta=0.20$, $s=0.5$ and the much 
smaller angular momentum loss rate of gravitational radiation. Thus 
these parameters mimic an unstable system just below the period gap. 
As we have seen at the end of Sect.~6.1, CVs just below the period 
gap can be unstable if the constant flux model with $s=0.5$ is 
adopted. They are, however, stable if the point source model is used, 
unless $\zeta_R$ is lowered below the standard value by invoking CAML 
(see MF98). If we use in addition to the constant flux model also the 
one-zone model in the limit $T_B \to T_1$, at low fluxes the donor 
star is even more susceptible to irradiation than if a more realistic 
form of $g(x)$ had been used. {\rev However,} for the purpose of this 
exercise it does not matter whether CVs below the gap are stable or 
not. This example was chosen just to demonstrate that with a low 
enough $\langle-\dm\rangle$ {\rev and depending on the value of 
$\eta$} a system {\rev could} evolve through mass transfer cycles as 
we have concluded from our stability discussion in Sects.~4 and 6. 
One additional property of this run which is worth mentioning is that 
the cycles are only rather weakly damped. The reason for this is that
a star with a mass $M_s < 0.25\msol$ is always fully convective and 
therefore ${\cal F}=7/3$ remains constant. Thus the decrease of 
$\zeta_\irr$ is mainly due to the slow increase of $\tau_{\rm KH}$ with 
time.

\begin{figure}[t] 
\plotone{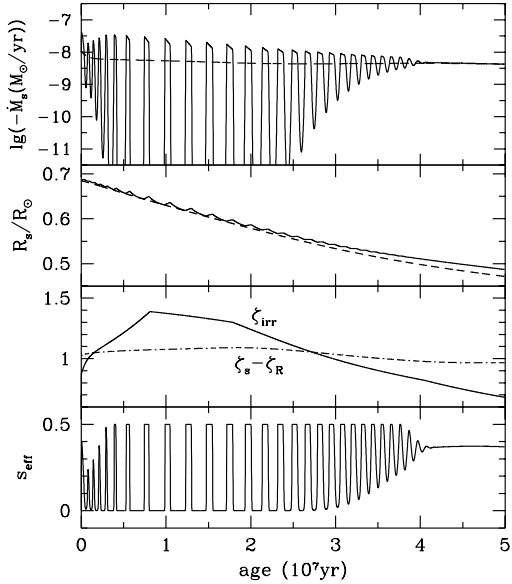}
\caption[]{As Fig. 11, but with an initial secondary mass
$M_{s,i}=0.7\msol$. Further parameters and model assumptions are listed
in Table 2. }
\end{figure}

\begin{figure}[t] 
\plotone{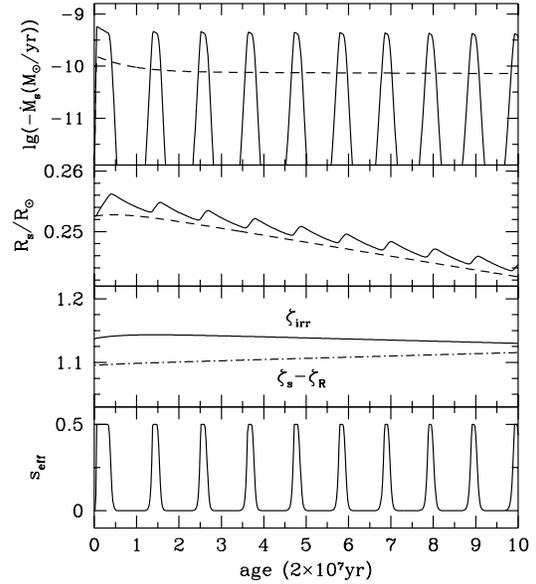}
\caption[]{As Fig. 11, but with an initial secondary mass 
$M_{s,i}=0.25\msol$ and assuming orbital angular momentum loss only
via gravitational radiation. Further parameters and model assumptions 
are listed in Table 2. }
\end{figure}

\section{Discussion and conclusions}

With the exception of the numerical examples presented in the 
previous section, where we have assumed the donor star to be on the 
ZAMS, we have so far not made any explicit assumption about the 
evolutionary status of the secondary. In fact our stability 
considerations are valid for any type of donor star as long as it 
has a sufficiently deep outer convective envelope. In the following 
we shall therefore briefly address the stability of mass transfer 
and the peak mass transfer rate in case of 
instability if the secondary is {\rev more or less nuclear evolved. 
In addition we shall briefly discuss the impact of the donor's 
metallicity on the stability of mass transfer.} A detailed 
investigation of the irradiation instability in those cases is, 
however, deferred to a subsequent paper.

\subsection{{\rev Effects of evolution and metallicity on the} 
stability of mass transfer}
Returning to the stability criterion (\ref{eq41}) we see that the 
left-hand side $\Lambda$ depends essentially only on the ratio of 
two time scales, i.e. the thermal time scale of the convective 
envelope $\tau_{\rm ce}$ and the mass transfer time scale 
$\tau_{M_s}$ or, respectively, on the time scale on which mass 
transfer is driven $\tau_{\rm dr} = (2/\tau_J+1/\tau_\nuc)^{-1}$. 
{\rev Mass transfer is} more stable the larger the ratio 
$\tau_{\rm ce}/\tau_{\rm dr}$. 

Evolutionary computations such as those shown in Ritter (1994) 
show that at a given secondary mass $\tau_{\rm ce}$ {\rev is 
the shorter the more evolved the star. If magnetic braking 
according to Verbunt \& Zwaan (1981) is invoked for computing the 
angular momentum loss rate, numerical computations (e.g. those 
shown in Ritter 1994, or King, Kolb \& Burderi 1996) of the mass 
transfer from a nuclear evolved donor show that 
$\tau_{\rm dr}$ is the longer the 
more evolved the donor at the {\em onset} of mass transfer. Taken 
together this means that mass transfer is the more unstable (less 
stable) the more evolved the donor. As an example, evaluating} 
$\Lambda$ along the evolutionary tracks shown in Ritter (1994) we 
find that as long as the secondary has not reached the terminal 
age main sequence, mass transfer is stable if $P\la 5^h-6^h$. In 
the sequence starting with a $1.2\msol$ {\rev secondary at the end 
of central hydrogen burning}, systems down to an orbital period 
$P \approx 2^h$ may be unstable.

{\rev Main sequence stars of low metallicity are systematically 
hotter, smaller, more luminous and have a thinner convective 
envelope than main sequence stars of solar metallicity. Hence 
$\tau_{\rm ce}$ is systematically shorter for low metallicity 
stars. At the same time $\tau_{\rm M_{\rm s}}$ is the shorter the 
lower the metallicity. Taken together this means that mass transfer 
is systematically more unstable (less stable) the lower the 
metallicity. As an example, using results from Stehle et al. (1997), 
we find that in CVs with a Pop. II donor mass transfer could be 
unstable for orbital periods as low as $\sim 4^h$.}
 
{\rev The case of} systems containing a giant donor and in 
which mass transfer is driven by its nuclear evolution, i.e. 
$\tau_{\rm dr} = \tau_\nuc$, has already been dealt with 
in some detail by KFKR97. We shall mention here only that 
such systems are systematically more unstable against 
irradiation-induced runaway mass transfer {\rev than systems with 
a main sequence donor. There are three 
reasons for this: First, for a giant $\tau_{\rm ce}$ is 
systematically shorter than for a main sequence star of the same
mass (because of the much larger values of $R$ and $L$), and this
despite the fact that $M_{\rm ce}/M$ can be quite large. Second, in
terms of radius changes due to irradiation, giants react much more
strongly than main sequence stars. (Note that in the framework of
the homology model presented in Sect. 3 the effective $\nu$ for 
giants is -3!) Third, the mass transfer time scale 
$\tau_{\rm M_{\rm s}}$associated with nuclear evolution of a giant 
is always much longer than $\tau_{\rm ce}$ over the entire range of 
core masses of interest, i.e. $0.15 \msol \la M_{\rm c} \la 0.5 \msol$. 
For details see KFKR97 and Ritter (1999).} 

\subsection{The peak mass transfer rate}
Our estimate (\ref{eq74}) of the peak mass transfer rate achieved 
during a mass transfer cycle can be rewritten as 
\be
{\rm Max}(-\dm) \approx \langle -\dm \rangle 
\left[1+\frac{2s}{\Lambda}\right] \quad.
\label{eq76}
\ee
In the case of a fully developed instability it is the second term in
(\ref{eq76}) which dominates. Therefore, the peak mass transfer rate
is essentially determined by the rate of expansion of the donor, i.e.
\be
\label{eq77}
{\rm Max}(-\dm) \approx \langle-\dm \rangle \frac{2s}{\Lambda}
\approx \frac{M_s}{\zeta_S-\zeta_R}\frac{s}{\tau_{\rm ce}} \quad.
\ee
Thus the lower the thermal inertia of the convective envelope the 
higher the peak mass transfer rate. As a consequence, very high peak 
rates can {\rev result} if the mass of the convective envelope or 
$\tau_{\rm KH}$ is small. {\rev If this is the case the long-term 
evolution of the corresponding systems could be drastically changed. 
We note e.g. that even if the donor is a main sequence star}, the 
peak mass transfer rate can easily exceed the value required for 
maintaining stable nuclear burning on the white dwarf, i.e. 
$\dot{M}_{\rm WD}\ga 10^{-7}\msol {\rm yr}^{-1}$ (e.g. Fujimoto
1982). With a Pop. I ZAMS donor this limit is reached if 
$M_s\ga 0.8\msol$ (see Fig. 11), with a donor at the end of central 
hydrogen burning if $P\ga 10^h$ or $M_s\ga 0.6 \msol$. These values 
are derived from the evolutionary calculations discussed in Ritter 
(1994).

The consequences of reaching peak mass transfer rates equal to or 
even in excess of the stable nuclear burning limit of the white 
dwarf can be far-reaching. First, if steady nuclear burning on the 
white dwarf is reached, such a system will no longer appear as an 
ordinary CV but rather look more like a supersoft X-ray source. 
Because such systems are bright in the EUV only, they become 
virtually undetectable in our Galaxy: very few such systems have 
{\rev indeed} been found (see e.g. Greiner 1996). Second, with 
nuclear burning on the white dwarf, the nuclear luminosity exceeds 
the accretion luminosity by a factor of $\sim 10-10^2$. With so 
much more irradiation luminosity available effects other than those 
discussed in this paper might also become important, e.g. driving 
of a strong wind from the donor (see van Teeseling and King 1998, 
King and van Teeseling 1998), and which would change the evolution 
of such systems altogether. Third, {\rev an unavoidable consequence 
of} the very high peak mass transfer rates {\rev are} very extended
low states during which a system is practically detached and thus 
virtually undetectable. Fourth, from the fact that such systems 
are barely detectable in both the high and low state and that the 
transition time between high and low states and vice versa is very 
short (see KFKR97), one is practically forced to the conclusion 
that the observed long-period CVs are either stable against the 
irradiation instability or, for some other reason do not reach peak 
rates equal to or larger than the stable nuclear burning limit.

\subsection{Conclusions}
In this paper we have studied the reaction of low-mass stars to
anisotropic irradiation and its implications for the long-term 
evolution of compact binaries. For this we have shown that the 
case of anisotropic irradiation in close binaries is relevant 
and that spherically symmetric irradiation is probably not an 
adequate approximation. We have studied the reaction of low-mass 
stars to anisotropic irradiation by means of simple homology 
considerations. We have shown in the framework of this model 
that if the energy outflow through the surface layers of a 
low-mass {\rev main sequence} star is blocked over a fraction
$s_\eff<1$ of its surface, it will inflate only modestly, by 
about a factor $\sim (1-s_\eff)^{-0.1}$ (see Eq. \ref{eq17}) in 
reaching a new thermal equilibrium, and that the maximum 
contribution to mass transfer due to thermal relaxation is 
$s_\eff$ times the value one obtains for spherically symmetric 
irradiation (i.e. $s_\eff=1$). The duration of the thermal 
relaxation phase is $0.1\tau_{\rm ce}|\ln(1-s_\eff)|$ (Eqs. 
\ref{eq22} and \ref{eq23}), where $\tau_{\rm ce}$ is the thermal
time scale of the convective envelope. In addition, we have 
carried out numerical computations of the thermal relaxation 
process of low-mass stars involving full   stellar models and 
using the modified Stefan-Boltzmann law (\ref{eq9}) as an outer 
boundary condition. The results of these computations (shown in 
Figs. 2-4) fully confirm results from homology and show that the 
effects caused by anisotropic irradiation are not only 
quantitatively but also qualitatively different from those caused 
by spherically symmetric irradiation.

Next we have carried out a detailed stability analysis. The 
criterion for stability against irradiation-induced runaway mass 
transfer in its most general form (arbitary irradiation geometry) 
is given in Eqs. (\ref{eq41}) and (\ref{eq46}). One of the 
remarkable results of this stability investigation is that 
it is not arbitrarily large irradiation fluxes which destabilize a 
system most effectively. Rather the most effective irradiating flux 
is $F_\irr \approx F_0$, where $F_0$ is the surface flux of the 
unperturbed star.

The reaction of the stellar surface to irradiation is best expressed 
in terms of a function $g(x) = -dF/dF_\irr$, where $x=F_\irr/F_0$ is 
the normalized flux. General considerations show that 
${\rm Max}(xg(x))<1$. For determining $g(x)$ we used a simple, 
analytic one-zone model for the superadiabatic convection zone. The 
results of this simple model (Eq. \ref{eq71}) are found to be in 
good qualitative and satisfactory quantitative agreement with 
results of computations involving full stellar models (see Fig. 8).

Application of our stability analysis to CVs and LMXBs yields the
following results:

CVs which evolve according to the standard evolutionary paradigm, 
i.e. the model of disrupted magnetic braking, are stable against
irradiation-induced runaway mass transfer if the mass of the 
(ZAMS) donor is $M_s\la 0.7\msol$. This holds unless substantial 
consequential angular momentum losses greatly destabilize the 
systems. Systems in which the mass of the (ZAMS) secondary is 
$M_s\ga 0.7\msol$ can be unstable but need not be so, depending on 
the efficiency of irradiation $\eta$ (defined in Eq. (\ref{eq24})).
Substantial consequential angular momentum losses can however
destabilize CVs over essentially the whole range of periods of 
interest.

CVs with a Pop. II or an evolved secondary are inherently less 
stable than CVs with {\rev a} Pop. I ZAMS secondary. The latter 
are those stars which, for a given mass, have the highest thermal 
inertia, making the corresponding CVs the most stable ones.

If mass transfer is unstable, we found that it must evolve through 
a limit cycle in which phases of irradiation-induced mass transfer 
alternate with phases of small (or no) mass transfer. At the peak 
of a cycle mass transfer proceeds on a time scale which is roughly 
$1/s_\eff$ times the thermal time scale of the convective envelope 
(see Eq. \ref{eq77}). With decreasing mass of the secondary the 
amplitude of the mass transfer cycles gets smaller and the cycles 
eventually disappear (after a system has become stable) because 
the thermal inertia of the secondary (expressed in the functions 
$\Lambda$ and $\Gamma$ defined respectively in Eqs. (\ref{eq41}) 
and (\ref{eq46})) increases strongly (see Figs. 10 and 11). A 
necessary condition for the maintenance of cycles is that the 
thermal time scale of the convective envelope has to be much 
shorter ($\la 0.05$) than the time scale on which mass transfer 
is driven.

LMXBs are very likely to be stable because a) the donor star is 
strongly shielded from direct irradiation over most of the 
hemisphere facing the X-ray source, and b) because where this is 
not the case, $xg(x)$, i.e. the sensitivity of the stellar surface 
to changes in the irradiating flux, is very small.

\acknowledgements{
We acknowledge support from a Royal Society/Chinese Academy of 
Sciences Joint Project. ZZ acknowledges support from Chinese 
National Natural Science Foundation. Major parts of this work 
have been completed while ZZ was visiting the MPA Garching, 
funded by the Max-Planck-Gesellschaft. We thank Andrew King 
for improving the language of the manuscript and an anonymous 
referee for helpful comments.}

\appendix
\section{Lateral heat diffusion} 
Here we show that in a realistic situation the lateral temperature 
gradient caused by irradiating the secondary anisotropically is 
negligible compared to the radial temperature gradient in the 
subphotospheric layers and that, therefore, lateral energy 
transport by radiation is unimportant. We demonstrate this in the 
framework of the point source model. In that model we estimate the 
lateral gradient of $T^4_\irr$ to be 
\bea
\label{a1}
(\nabla T^4_\irr)_\vart &\approx&
\frac{T^4_\irr (\vart=0) - T^4_\irr (\vart=\vart_\max)}
{R_s \vart_\max} \nonumber\\
 &=& \frac{T^4_\irr (\vart = 0) - T^4_0}{R_s \vart_\max} \quad.
\eea
On the other hand, the radial temperature stratification is given 
by the Eddington approximation (e.g. Tout et al. 1989) 
\bea
\label{a2}
T^4(\tau,\vart) &=& \frac{3}{4} \, 
\rund{T^4_\irr(\vart) - \frac{F_\irr(\vart)}{\sigma}}
 \; \rund{\tau+\frac{2}{3}} \nonumber\\
&+& \frac{F_\irr(\vart)}{\sigma} \quad,
\eea
where $\tau$ is the optical depth. Therefore 
\be
(\nabla T^4)_r = \frac{dT^4(\tau,\vart)}{dr} = \frac{3}{4} \, 
\rund{T^4_\irr(\vart) - \frac{F_\irr(\vart)}{\sigma}} \; 
\frac{d\tau}{dr}\label{a3}
\ee
and 
\be
\mbox{Min}\vert(\nabla T^4)_r\vert = \frac{3}{4} \, 
\rund{T^4_\irr(\vart=0) - \frac{F_\irr(\vart=0)}{\sigma}}\; 
\frac{d\tau}{dr}\quad.\label{a4}
\ee
From the definition of the optical depth we have 
\be
\frac{d\tau}{dr} = -\kappa\varrho\quad.\label{a5}
\ee
Inserting (\ref{a5}) into (\ref{a4}) yields the minimum radial 
temperature gradient in the presence of irradiation: 
\be
\mbox{Min} \vert(\nabla T^4)_r\vert = \frac{3}{4}\, 
\rund{T^4_\irr(\vart=0) - \frac{F_\irr (\vart = 0)}{\sigma}}\; 
\kappa\varrho\quad.\label{a6}
\ee
Combining now (\ref{a1}) and (\ref{a6}) we obtain 
\bea
\label{a7}
&&\mbox{Max}\vert\frac{(\nabla T^4)_\vart}{(\nabla T^4)_r}\vert 
\approx \nonumber\\
&& \frac{4}{3\vart_\max} \; 
\frac{T^4_\irr (\vart=0) - T^4_0}{T^4_\irr (\vart = 0) - 
F_\irr(\vart=0)/\sigma} \; \frac{1}{\kappa \varrho R_s}\quad.
\eea
Now, because $4/3 \vart_\max \approx 1$ and in the subphotospheric 
layers $\kappa \varrho H_P \approx 1$, we have 
\bea
\label{a8}
\mbox{Max}\vert\frac{(\nabla T^4)_\vart}{(\nabla T^4)_r}\vert &\approx& 
\frac{H_P}{R_s}\; \frac{T^4_\irr(\vart = 0) - T^4_0}
{T^4_\irr (\vart = 0) - \frac{F_\irr (\vart=0)}{\sigma}} \nonumber\\
&=& \frac{H_P}{R_s} \; 
\frac{G(x(\vart=0)) + x(\vart=0)-1}{G(x(\vart=0))}\quad.
\eea
Because in the stars in question $H_P/R_s\approx 10^{-4}$ is so 
small, the lateral temperature gradient is always much smaller 
than the radial one unless $x(\vart=0)\gg 1$ and thus 
$G(x(\vart=0))$ $\ll 1$. However, in the range of interest of 
$x$ and $G(x)$, i.e.\ where $xg(x)$ is near its maximum and 
$x\approx 1$ to a few, (\ref{a8}) yields that the lateral 
temperature gradient is of order 
$(\nabla T)_\vart \approx H_P/R_s (\nabla T)_r \ll (\nabla T)_r$.

\end{document}